\documentclass{aa}  
\usepackage{graphicx,natbib,psfig}  
\newcommand{\ltsima} {$\; \buildrel < \over \sim \;$}  
\newcommand{\gtsima} {$\; \buildrel > \over \sim \;$}  
\newcommand{\lta} {\lower.5ex\hbox{\ltsima}}  
\newcommand{\gta} {\lower.5ex\hbox{\gtsima}}

\begin{document}  

\title{On the nature of optical nuclei in FR~I radio-galaxies
from ACS/HST imaging polarimetry 
\thanks
{Based  on observations obtained at
the  Space  Telescope Science  Institute,  which  is  operated by  the
Association of  Universities for Research  in Astronomy, Incorporated,
under NASA contract NAS 5-26555.}}
  
\titlerunning{Polarimetry of FR~I radio-galaxies nuclei}  
\authorrunning{A. Capetti et al.}
  
\author{Alessandro Capetti
\inst{1}
\and  David J. Axon \inst{2}  
\and  Marco Chiaberge \inst{3}
\and  William B. Sparks \inst{3,4}
\and  F. Duccio Macchetto \inst{3,4}
\and  Misty Cracraft \inst{3}
\and  Annalisa Celotti \inst{5}}
   
\offprints{A. Capetti}  
     
\institute{INAF - Osservatorio Astronomico di Torino, Strada
  Osservatorio 20, I-10025 Pino Torinese, Italy\\
\email{capetti@to.astro.it}
\and 
Department of Physics,
Rochester Institute of Technology, 85 Lomb Memorial Drive,
Rochester, NY 14623, USA
\and
Space Telescope Science Institute
3700 San Martin Drive, Baltimore, MD 21218, USA
\and
Affiliated with  ESA's Research \& Space Science Department
\and
SISSA/ISAS, Via Beirut 2-4, I-34014 Trieste, Italy}

\date{}  
   
\abstract{We obtained optical imaging polarimetry with the ACS/HRC
aboard the HST of the 9 closest radio-galaxies in the 3C catalogue with an FR~I
morphology. 
The nuclear sources seen in direct HST images in these galaxies
are found to be highly polarized with levels in the range $\sim$ 2 - 11
\% with a median value of 7 \%.
We discuss the different mechanisms that produce polarized emission
and conclude that the only viable interpretation is a 
synchrotron origin for the optical nuclei. 
This idea is strengthened by the analogy with
the polarization properties of BL Lac objects,
providing also further support to the FR~I/BL Lac unified model.
This confirms previous suggestions that the dominant emission mechanism
in low luminosity radio-loud AGN is related to non-thermal
radiation produced by the base of their jets.

In addition to the nuclear polarization (and  
to the large scale optical jets), polarization is detected 
co-spatially with the dusty circumnuclear disks, likely 
due to dichroic transmission; 
the polarization vectors are tangential to the disks 
as expected when the magnetic
field responsible for the grains alignment is stretched by differential
rotation. 

We explored the possibility to detect the polarimetric signature 
of a misaligned radiation beam in FR~I, expected in our sources in the frame of
the FR~I/ BL Lac unification. We did not find this effect in any
of the galaxies, but our the results are not conclusive on whether 
a misaligned beam is indeed present in FR~I.

 } \maketitle
  
\section{Introduction}
\label{introduction}

The presence of a radio-source represents a common manifestation of
nuclear activity associated with elliptical galaxies, in particular for
the brightest members of this class.  For example, among galaxies
brighter than $M_B < -21$, more than 20 \% are radio-active with
luminosities $L_{408 MHz} > 10^{23.5}$ W Hz$^{-1}$ \citep{colla75}. 
Most of them are low luminosity radio-galaxies (LLRG) and show the
characteristic edge-darkened FR~I radio morphology \citep{fanaroff74}. 

HST imaging of low luminosity radio-galaxies proved to be a 
very powerful tool for a better
understanding of the properties of this class of Active Galactic Nuclei (AGN) 
as it
allowed us to isolate, for the first time, their genuine optical nuclear
emission from that of the host galaxies.
Analysis of R band HST images of the sample
formed by all the 33 FR~I sources belonging to the 3C catalogue
showed that unresolved optical nuclear sources are
detected in the great majority of these galaxies
\citep{chiaberge:ccc}. 
Their optical and radio nuclear luminosities 
show a clear correlation, suggestive of a
common non-thermal synchrotron origin, most likely from the base of their
relativistic jets. 

Recently we found a similarly tight connection 
between the radio and X-ray luminosities measured from
Chandra observations \citep{balmaverde06}, confirming the trends 
already suggested by lower resolution X-ray telescopes
\citep[e.g.][]{fr1sed,hardcastle00,trussoni03,hardcastle03}. 
Furthermore, the correlation between radio, optical
and X-ray emission extends to the large population of radio-loud nuclei 
found in early-type galaxies, with luminosities smaller by a factor
as large as 1000 with respect to classical low-luminosity
radio-galaxies \citep{paper2}.

The picture which emerges is that the nuclear emission in low
luminosity radio-loud AGN is dominated by non-thermal emission from
the base of their jets. 

This result also provides strong support for the identification of 
FR~I radio-galaxies as the misoriented population
associated with BL Lac objects. Indeed, in the framework of this
unifying scheme, the non-thermal radiation from the jet, which
dominates the optical emission in BL Lacs, is expected to be present
also in FR~I radio-galaxies, although de-amplified.  

Alternative explanations for the origin of the nuclear sources
(e.g. stellar nuclear cusps, emission from radiatively efficient
accretion disks) appear to be less likely, principally as none of them
naturally accounts for the tight 
correlation between the radio, optical and X-ray 
luminosities. Apparently, the only viable
alternative are radiatively  inefficient accretion flows 
\citep[RIAF,][]{narayan95}.
Indeed, the measured nuclear optical luminosities translate
for a $10^9 M_{\sun}$ black hole as typical
of FR~I sources, into a fraction as small as 
$\la 10^{-5} - 10^{-7}$ of the
Eddington luminosity. This indicates that accretion occurs at low
rate and/or in a low radiative regime.

Polarimetric observations, particularly those at radio and optical wavelengths,
played an important role in extragalactic astrophysics. From the discovery of 
synchrotron radiation to the first good evidence for AGN unification, 
to the discovery of variable linear polarization in the absorption troughs 
of BAL QSO, polarization studies often provide the best and sometimes the 
only clues we have regarding the morphology and the radiation processes 
in these diverse sources. This is  also clearly the case for understanding
the origin of the nuclear emission in radio-galaxies.

In fact, a simple and direct test of the scenario presented above can be 
performed by measuring the {\sl nuclear} polarization of FR~I radio-galaxies.
If their nuclear emission is dominated by synchrotron radiation
we expect to detect significant polarization.

Ground based measurements \citep[see e.g. ][]{impey91} have 
been hampered by the dominant dilution of unpolarized starlight 
as their nuclear sources only account for a few per cent 
of the total flux in typical ground based apertures.
Thanks to the high resolution of HST it is possible to isolate 
the nuclear emission from that of the host galaxy and to measure
its intrinsic polarization.

Previously we obtained polarization measurements with HST
for two radio-galaxies. For the first galaxy, M~87 (AKA 3C~274) 
the nucleus was found to be only marginally 
polarized (P = 1.6 $\pm$ 0.7\%) in the optical 
\citep{capetti:m87}. Conversely, the nucleus of the second
radio-galaxy, Centaurus A, is highly polarized, 
with P = 11.1 $\pm$ 0.2\% at 2 $\mu$m \citep{capetti:cena}.
This high polarization can be interpreted in two different scenarii.
The first possibility is that we are seeing the infrared counterpart
of the synchrotron radio-core and this accounts naturally for its
polarization. However, since
the orientation of the nuclear polarization (PA = 148.2 $\pm$ 1.0) is almost
exactly perpendicular to the radio jet (PA = 55), the polarized
component might also be a very compact scattering region (with radius
smaller than 1 pc) and in this case the nucleus is hidden to our view.
We are thus left with a substantial ambiguity in interpreting the observed
nuclear polarization based on observations of a single object.

Only with measurements of the nuclear polarization
for a complete sample of objects it will be possible to settle this issue.
It is also essential that the sample is unbiased with respect
to orientation, so that the objects can be considered to have their
jet axes randomly distributed with respect to the line of sight.
The 3C sample is selected at low radio frequency 
where the extended emission, which does not depend on orientation,
dominates over the nucleus; thus the flux threshold of 
the catalogue selection criterion does not introduce any orientation bias.
We then selected, from the complete sample of the 33 FR~I part of the
3C sample, the closest sources setting a redshift 
threshold at z $<$ 0.025. This lead to the final sample of 9 sources.
In all of them an optical nuclear source has been detected
by \citet{chiaberge:ccc}.

The paper  is organized as  follows: in Sect.  \ref{observations} we
describe the observations of the 9 science targets. 
The results of Appendix \ref{calibration}, where
we discuss the capabilities of the HRC as a polarimeter, 
are used in Sect. \ref{nuclei} to measure the
polarization of the optical nuclei, while in Sect. \ref{off} we 
describe and discuss the nature of regions of extended
polarization. The origin of the nuclear emission is explored 
in Sect. \ref{science}. Finally in Sect. \ref{summary}
we discuss our results and present our conclusions.

\section{Observations and data reduction}
\label{observations}
Images of  the 9  radio-galaxies were obtained with the  High Resolution
Camera  (HRC) of  the Advanced  Camera  for Survey  (ACS). 
Its pixel size is 0\farcs028 $\times$ 0\farcs025, for a 
nominal 29\arcsec $\times$ 26\arcsec field of view. 
The broad-band filter F606W (centered at 5900  \AA\ and with a width
of 2300 \AA) was used in  combination with  the 
three visible  polarizers, POL0V,  POL60V and
POL120V. The polarizers have their principal planes oriented
nominally at PA 0, 60 and 120 degrees.  

Two exposures of equal length have been obtained for each filter combination 
to enable removal of cosmic ray events. 
The total exposures times per polarizer
range from 782 to 824 seconds for most objects,
while they are 300 s  for 
two sources (namely 3C~264 and 3C~274).
See Table \ref{log} for the observations log. 

The individual images were processed through the standard HST pipeline and
the exposures were combined to remove cosmic ray events with the 
{\sl multidrizzle}
task in the STSDAS {\sl dither} package. 

The  images obtained through the three polarizers
can  be  combined  to derive  the  Stokes
parameters I, Q and U using the following relationships

$$ I = { 2 \over 3} ( I_0 + I_{60} + I_{120} ) $$

$$ Q = { 2 \over 3} ( 2 \cdot I_0 - I_{60} - I_{120} ) $$

$$ U = { 2 \over {\sqrt 3}} ( I_{60} - I_{120} ) $$

\indent
where $I_0$, $I_{60}$  and $I_{120}$ are the count  rates measured with
the  three polarizers.   The Stokes  parameters  can be  then used  to
measure  the   polarized  flux  $I_P  =   (Q^{2}  +U^{2})^{1/2}$,  the
percentage of linear polarization $ P =  {I_P \over I} $ and the polarization
position angle (in the instrument reference frame) 
$ \theta_{\rm instr.} =0.5 \cdot tan^{-1} ({U \over Q})$. The polarization PA 
can be referred to the sky reference frame, following 
\citet{biretta04a,biretta04b}, as
$\theta =  \theta_{\rm instr.}$ -69.6 + PAV3, where PAV3 is the orientation
of the telescope V3 axis during the observations.

The statistical errors in the Stokes parameters were computed assuming
Poisson noise in the images.
The errors on the degree and position angle of
the polarization are then estimated as
$$\sigma_P = \sqrt{2} I_i^{-1/2} {\,\,\, \rm and \,\,\,} 
\sigma_{\theta} = \sqrt{2} I_i^{-1/2} P^{-1} $$
respectively, where $I_i$ are the total counts in one of the polarizers.
As explained in Appendix \ref{bias} 
we preferred not to correct for the polarization
bias.

In the Appendix \ref{calibration} we discuss the capabilities of the HRC as a polarimeter,
analyzing the complete set of observations of standard targets available from
the calibration program and assessing the accuracy level of polarization
measurements with this instrument.  Here we briefly summarize the main results
of this analysis and refer for the details to this Appendix.  

Measurements of
the count rates through each of the 3 polarizers for two unpolarized standard
stars indicate a substantial departure from an ideal polarimeter, for which
one would expect equal count rates.  We investigated the origin of this effect
by using also observations of a polarized star, observed in 5 different epochs
and at different telescope orientations.  Our conclusion is that the expected
polarization for all calibration targets can be accurately reproduced if there
is a difference in the transmission of the 3 polarizers, with POL0V being
about 10 \% less efficient than POL60V and POL120V.  By applying the
appropriate correction to the count rates we reproduced accurately the
expected values for the standard stars to within 0.4\%. This value can be
considered as upper limit to the uncertainties due to any residual
instrumental polarization, as it also includes errors due to the flat field
accuracy on the polarization measurements and to the
statistical uncertainties.  

This analysis was performed using relatively large
apertures (50 pixels) while for our science targets we are interested in
measuring the nuclear polarization on smaller apertures.  On a scale of a few
pixels, there might be e.g. differences in the PSF of each polarizer, leading
spurious polarization measurements.  We investigated this issue integrating
the count rates through several synthetic apertures and concluded that robust
polarimetric measurements can be obtained with apertures of radius
as small as r $\gta$ 3 pixels, with no appreciable differences in accuracy.

In the analysis of our science targets we therefore first applied the
appropriate correction to the count rates (dividing 
by 1.095 and 1.109 the POL60V and POL120V images respectively)
and only then we estimated the Stokes parameters.  


\begin{table}
\caption{Observations log.}
\label{log}
\centering
\begin{tabular}{l l c c c}
\hline\hline
Name & \multicolumn{2}{c}{\,\,\,\,\,\,\,\,\,\,\,\,\,\,\,\,\, Filter \,\,\,\,\,\,\,\,\,\,\,\,\,\,  Exp. time} & Date & PAV3 \\
\hline	       			      	 	                  
3C~31    & F606W+POLiV  & 796  &  02-08-25 &  50.6 \\
3C~66B   & F606W+POLiV  & 824  &  02-08-21 &  64.1 \\
3C~264   & F606W+POLiV  & 300  &  02-12-05 & 115.9 \\
3C~270   & F606W+POLiV  & 782  &  02-07-07 & 291.4 \\
3C~272.1 & F606W+POLiV  & 782  &  02-12-17 & 115.2 \\
3C~274   & F606W+POLiV  & 300  &  02-12-10 & 115.6 \\
3C~296   & F606W+POLiV  & 782  &  02-07-10 & 296.8 \\
3C~465   & F606W+POLiV  & 786  &  02-09-06 &  48.4 \\
3C~449   & F606W+POLiV  & 796  &  02-07-05 &  17.0 \\
\hline
\end{tabular}
\\ 
\end{table}

\section{Nuclear polarization measurements}
\label{nuclei}

\begin{figure*}
\centerline{
\psfig{figure=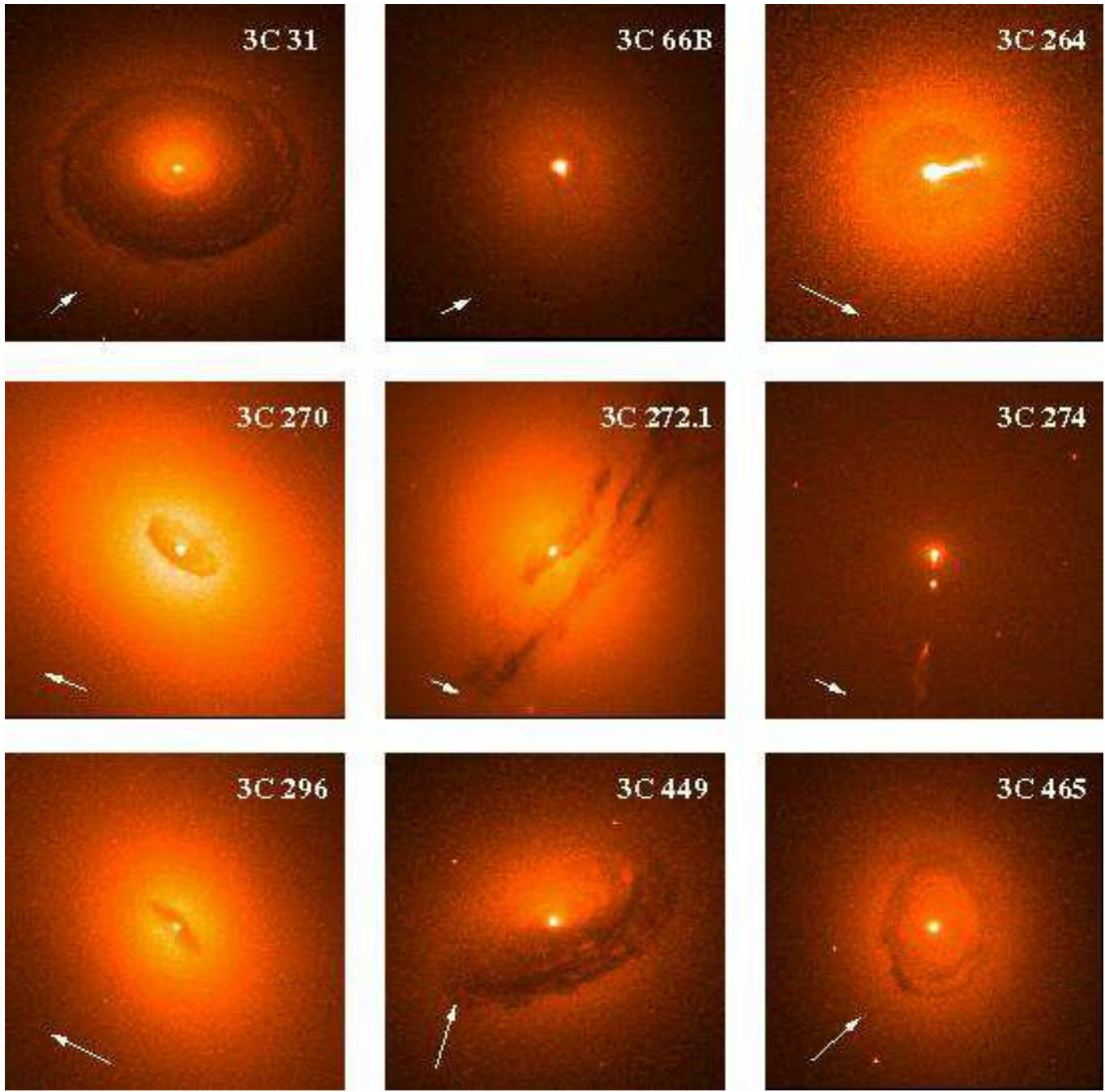,width=1.00\linewidth}
}
\caption{\label{images} ACS/HRC images of the 9 FR~I radio-galaxies. 
The arrow in the bottom left corner is 1\arcsec\ long and points toward North.}
\end{figure*}

In Fig. \ref{images} we show the images obtained with the F606W+POL0V
filter for our 9 targets. All these objects have been previously
observed at different wavelengths with HST with several instruments 
and their morphology is already well studied. 
We only point out the presence of the
optical nuclear sources, of which we aim to measure the polarization.
Furthermore, a circumnuclear dusty disk is found in every galaxy,
with the exception of 3C~274 where the gaseous disk is seen in
emission; the optical jets of 3C~264 and 3C~274 
already seen in previous HST images \citep{crane93,boksenberg92} are
also clearly detected.

\begin{figure*}
\centerline{
\psfig{figure=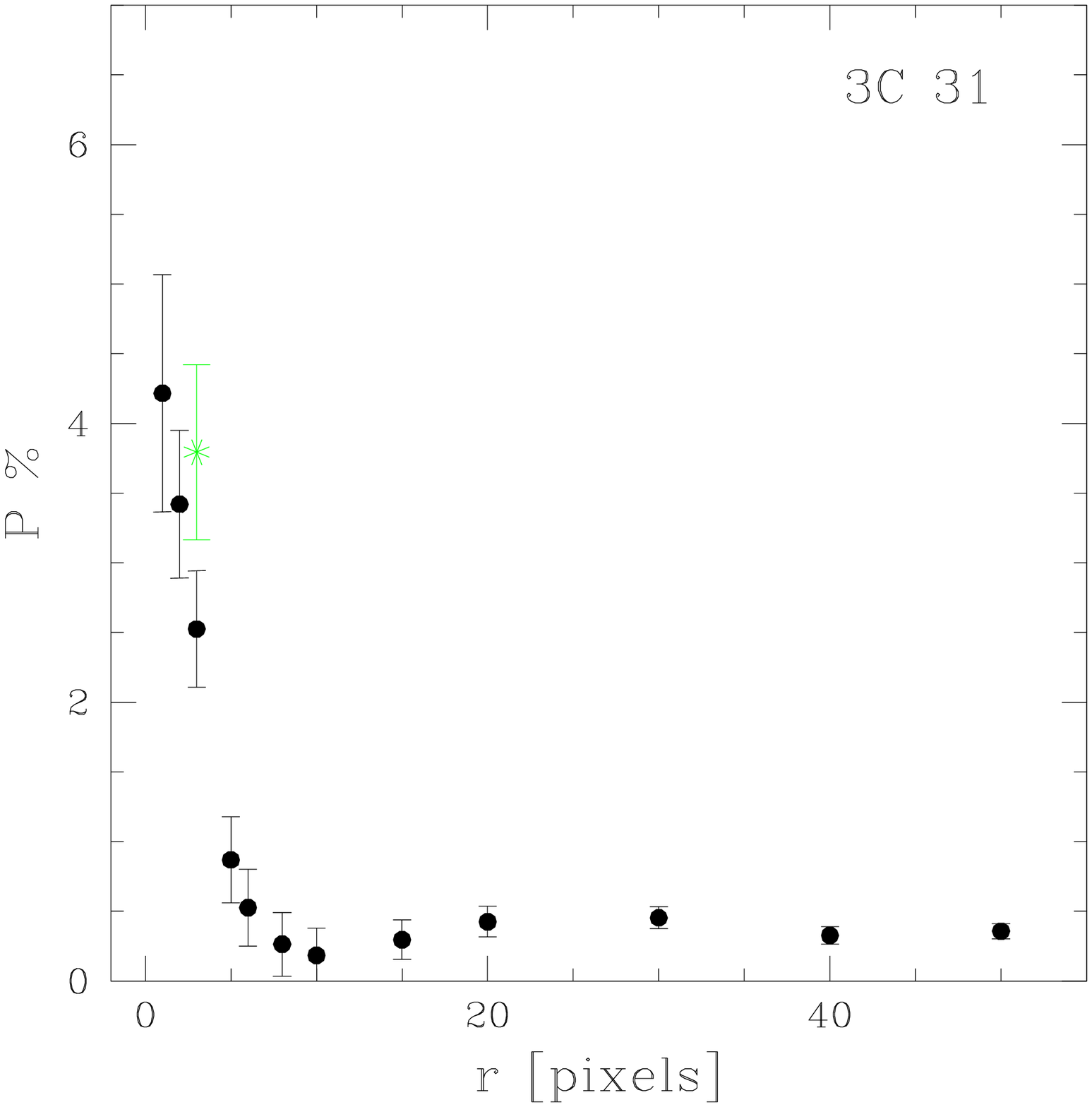,width=0.33\linewidth}
\psfig{figure=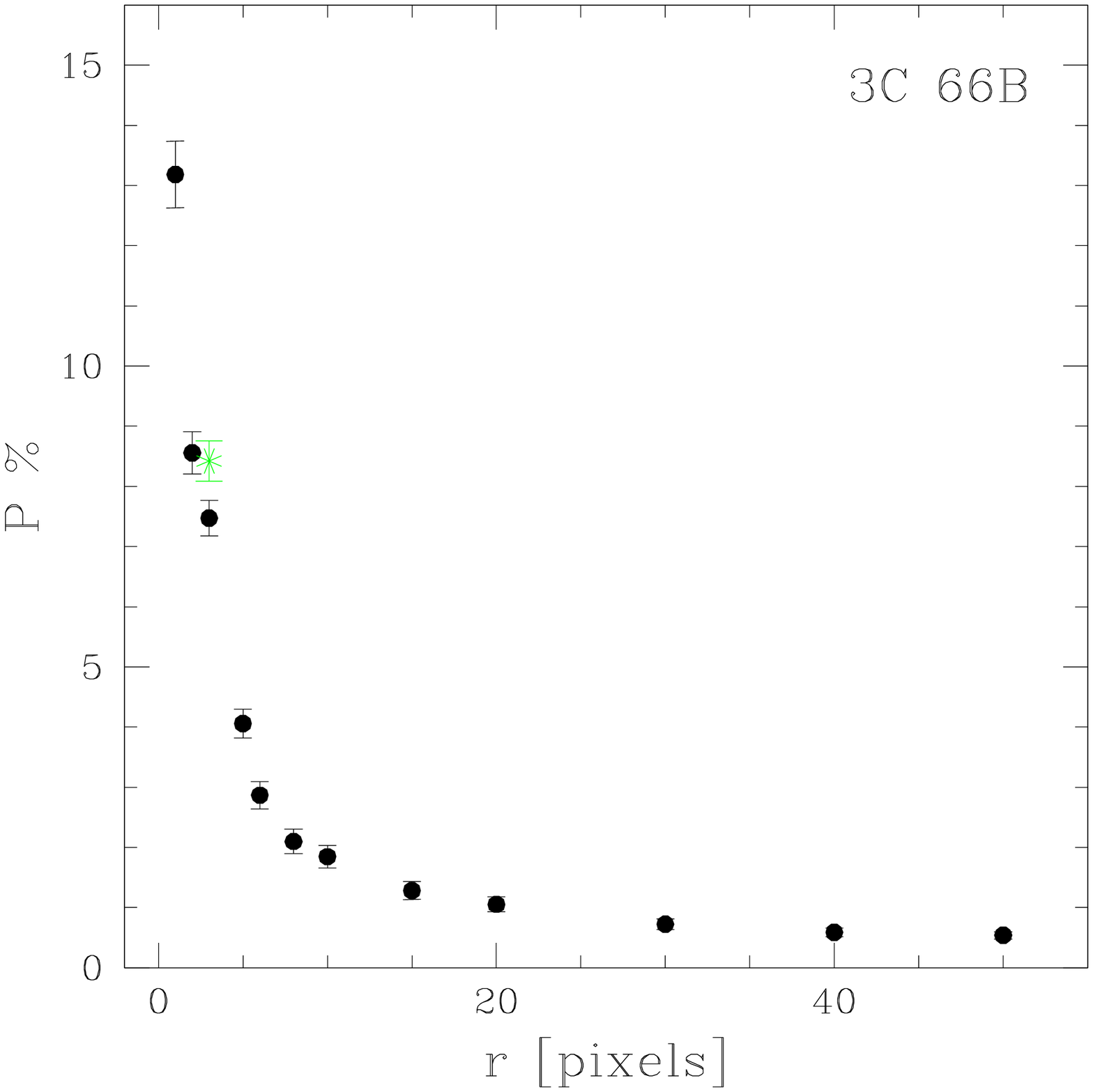,width=0.33\linewidth}
\psfig{figure=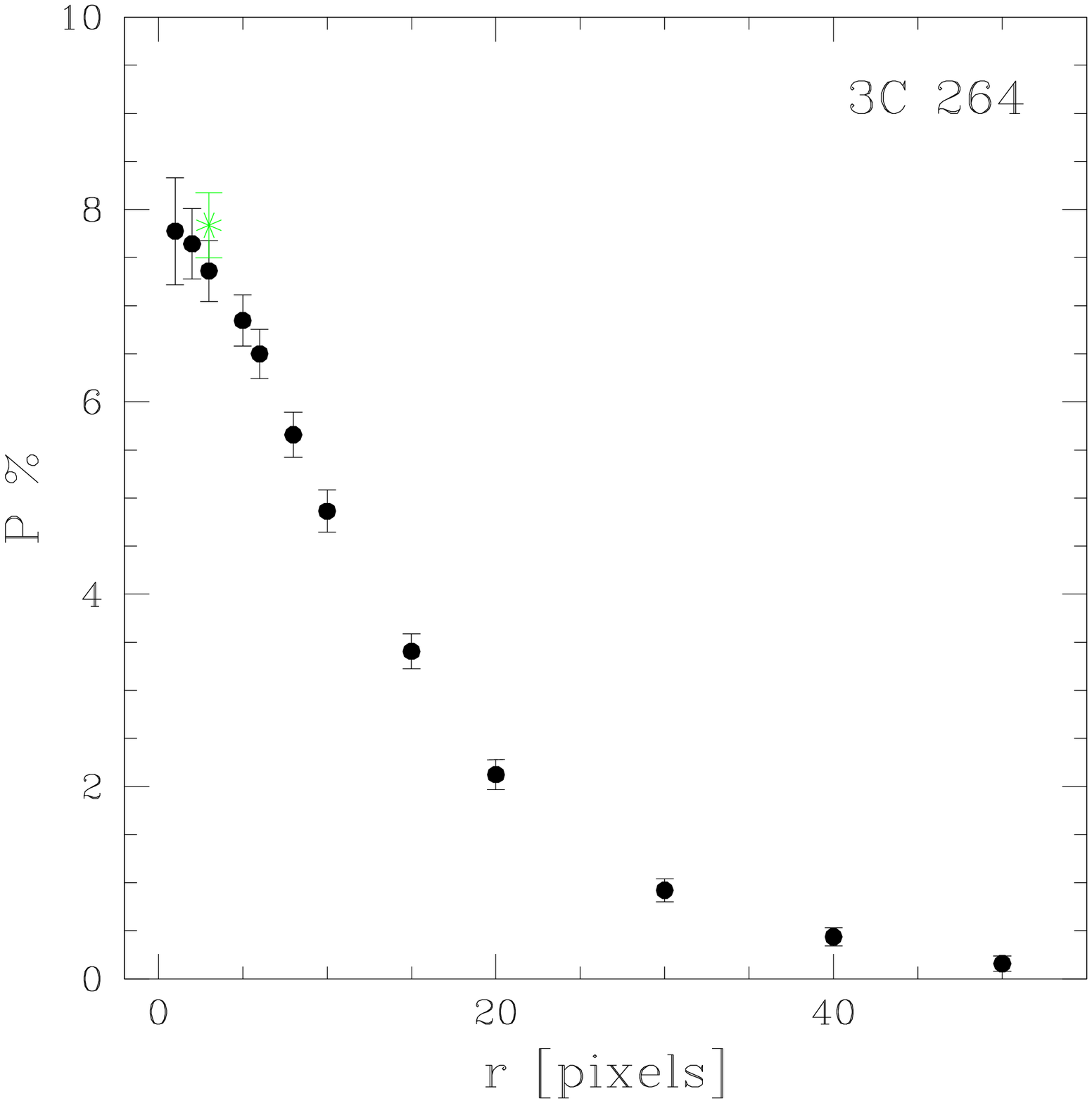,width=0.33\linewidth}}
\centerline{
\psfig{figure=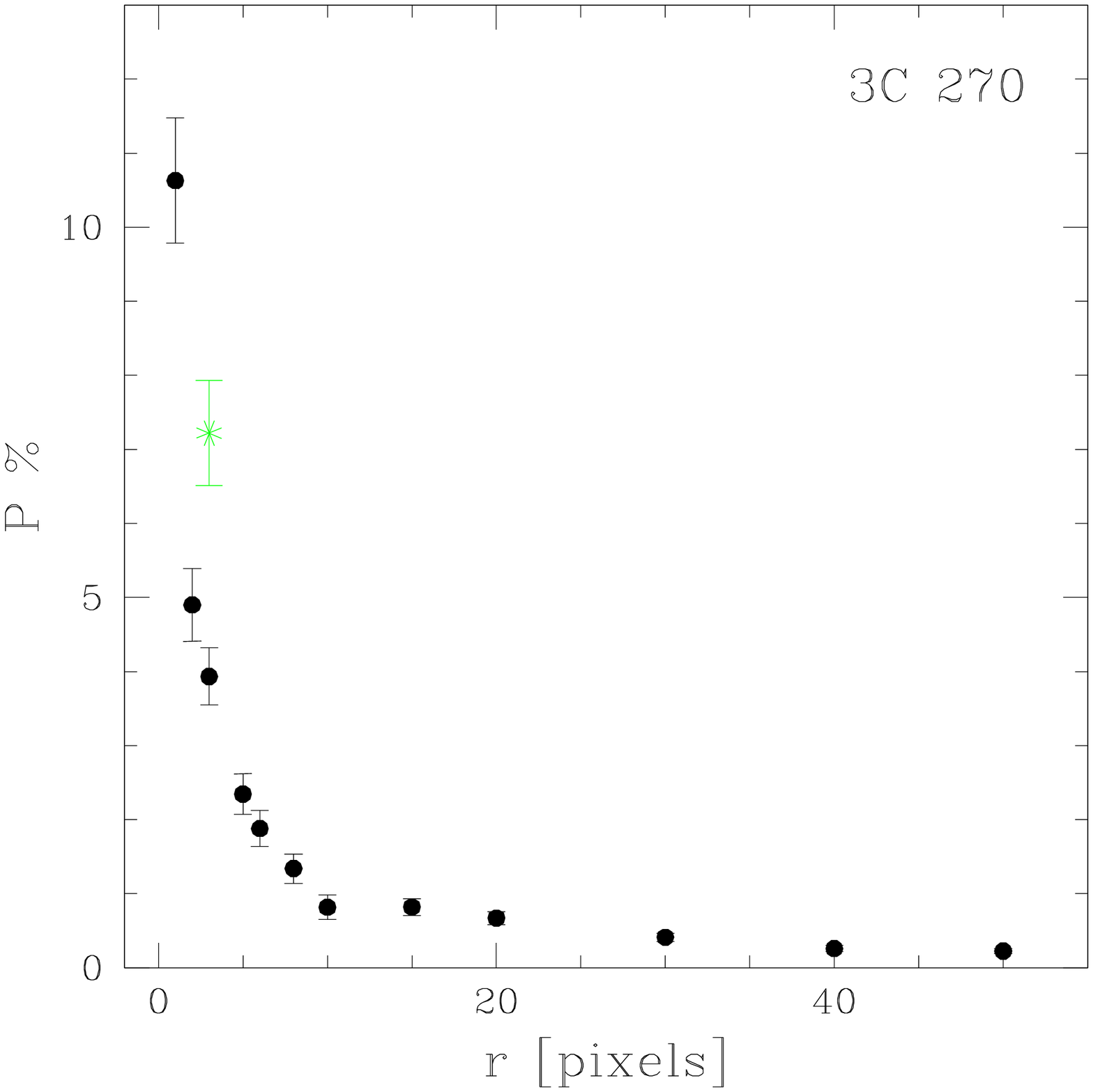,width=0.33\linewidth}
\psfig{figure=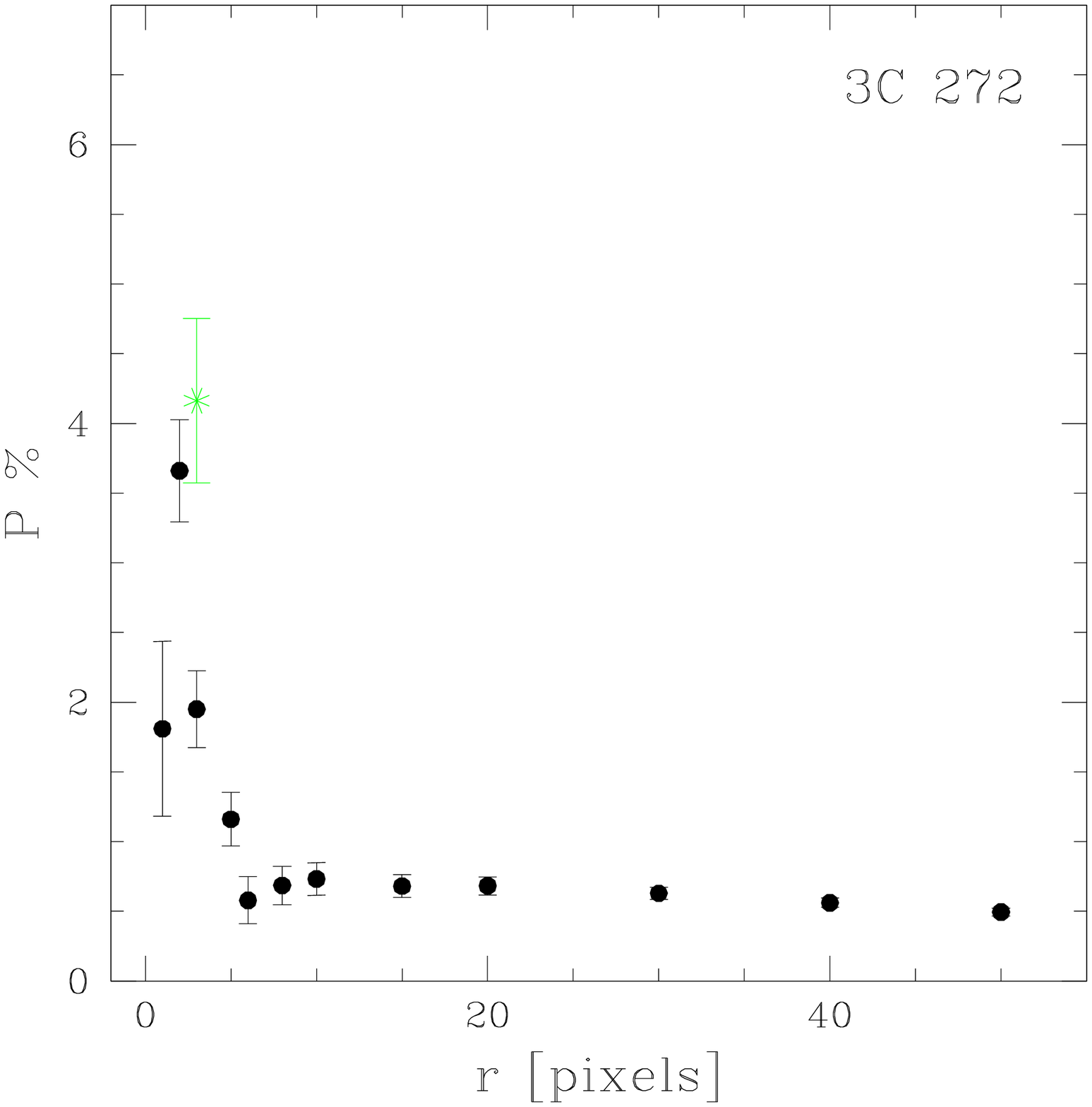,width=0.33\linewidth}
\psfig{figure=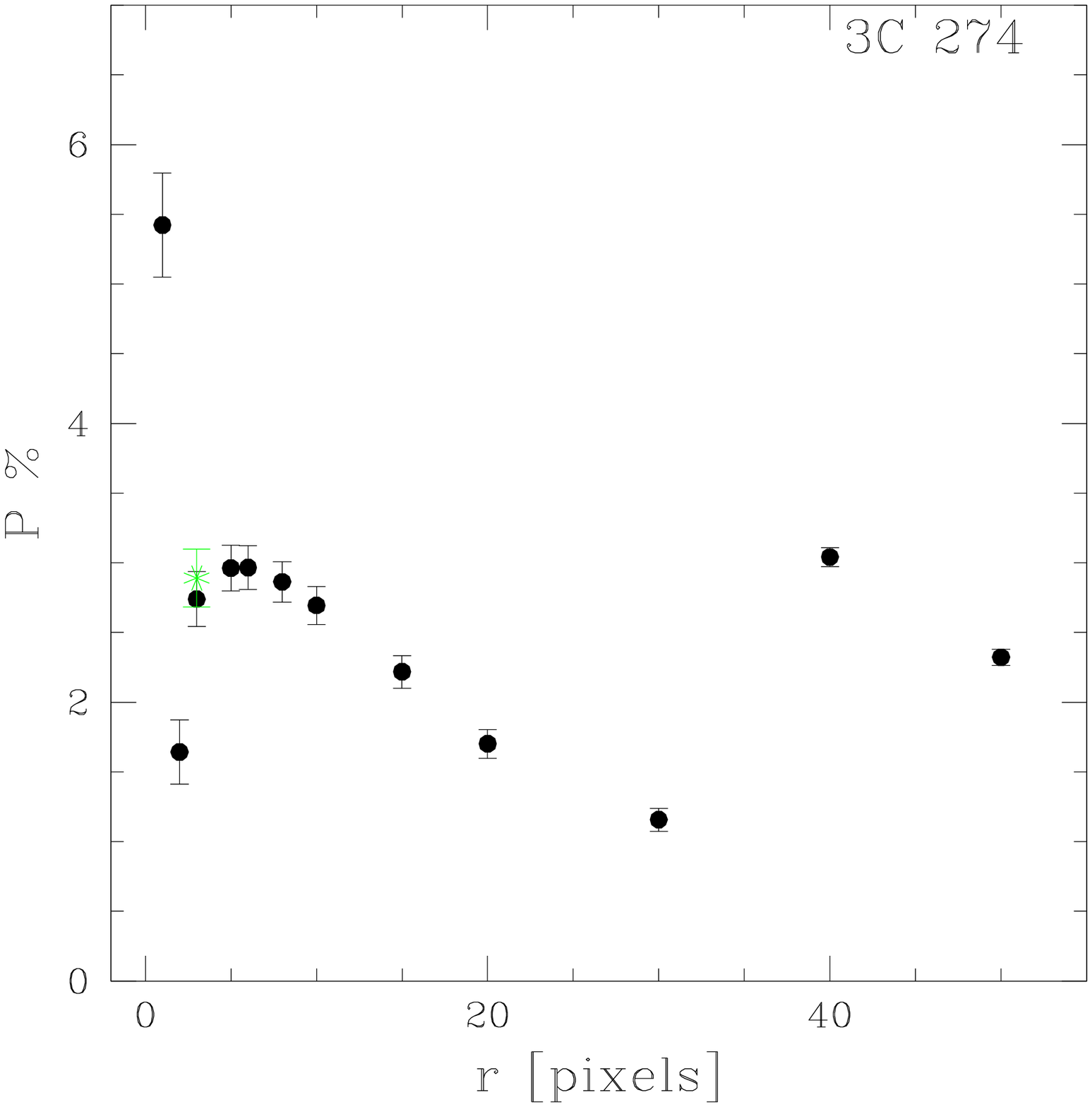,width=0.33\linewidth}}
\centerline{
\psfig{figure=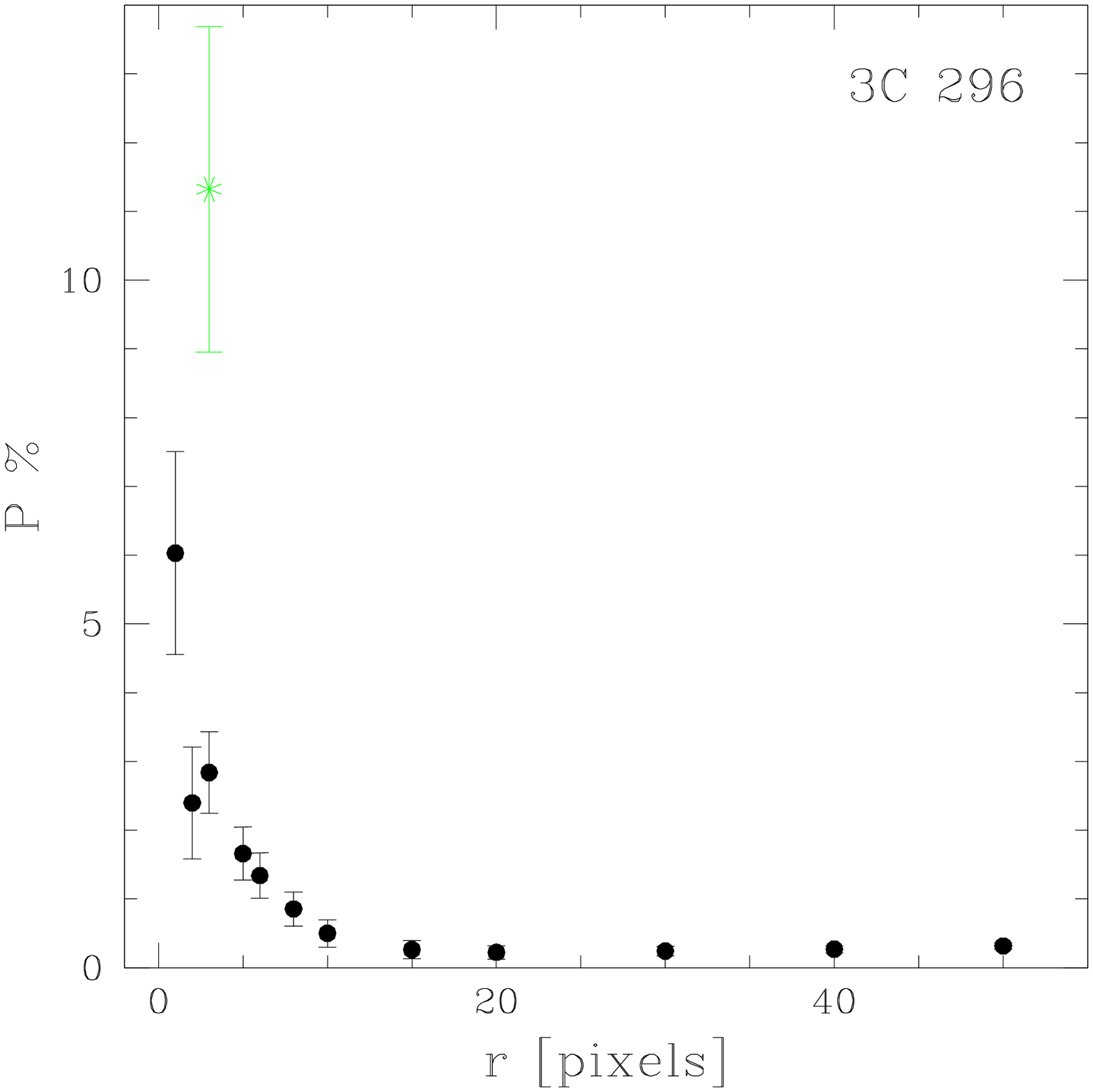,width=0.33\linewidth}
\psfig{figure=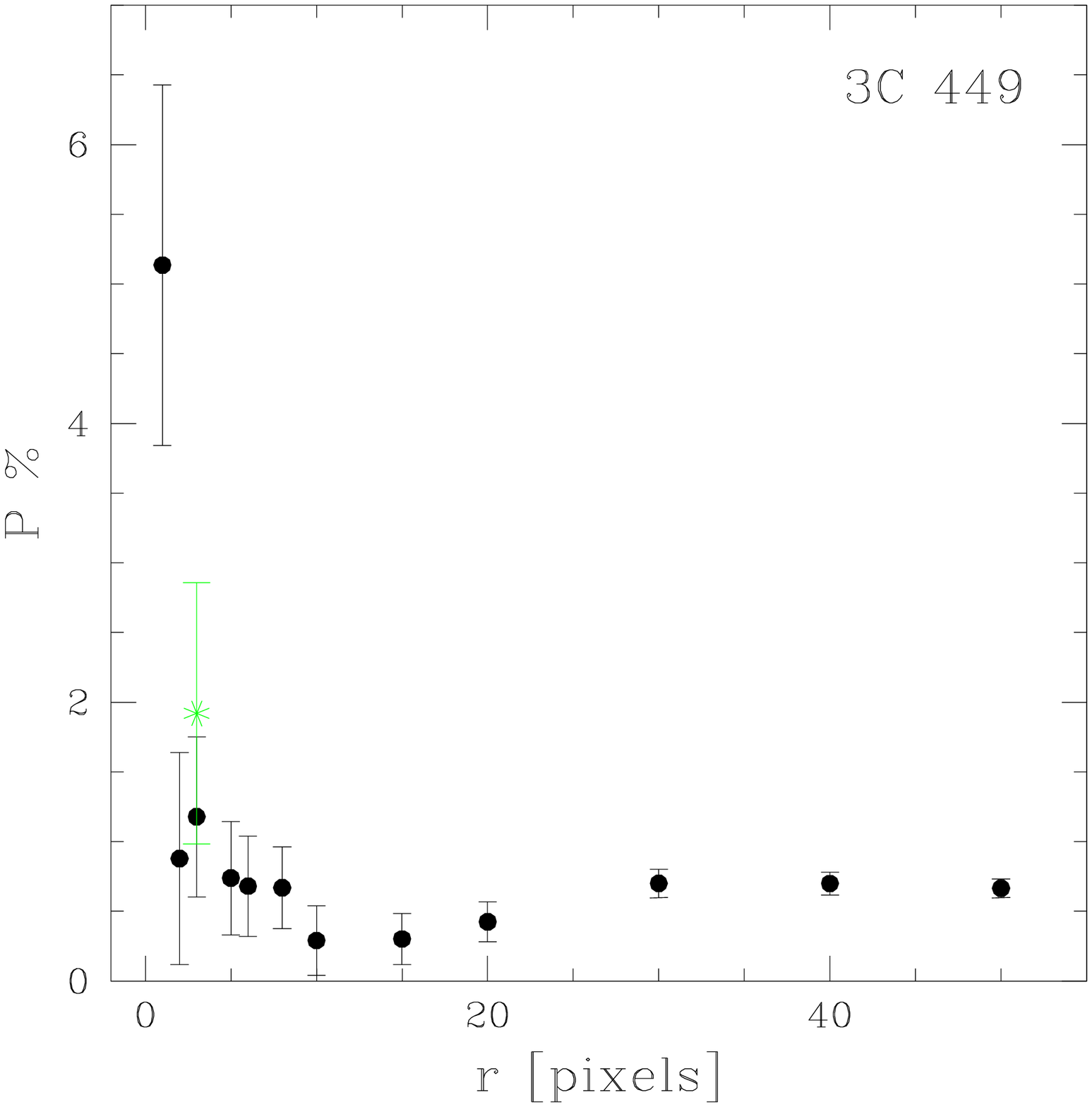,width=0.33\linewidth}
\psfig{figure=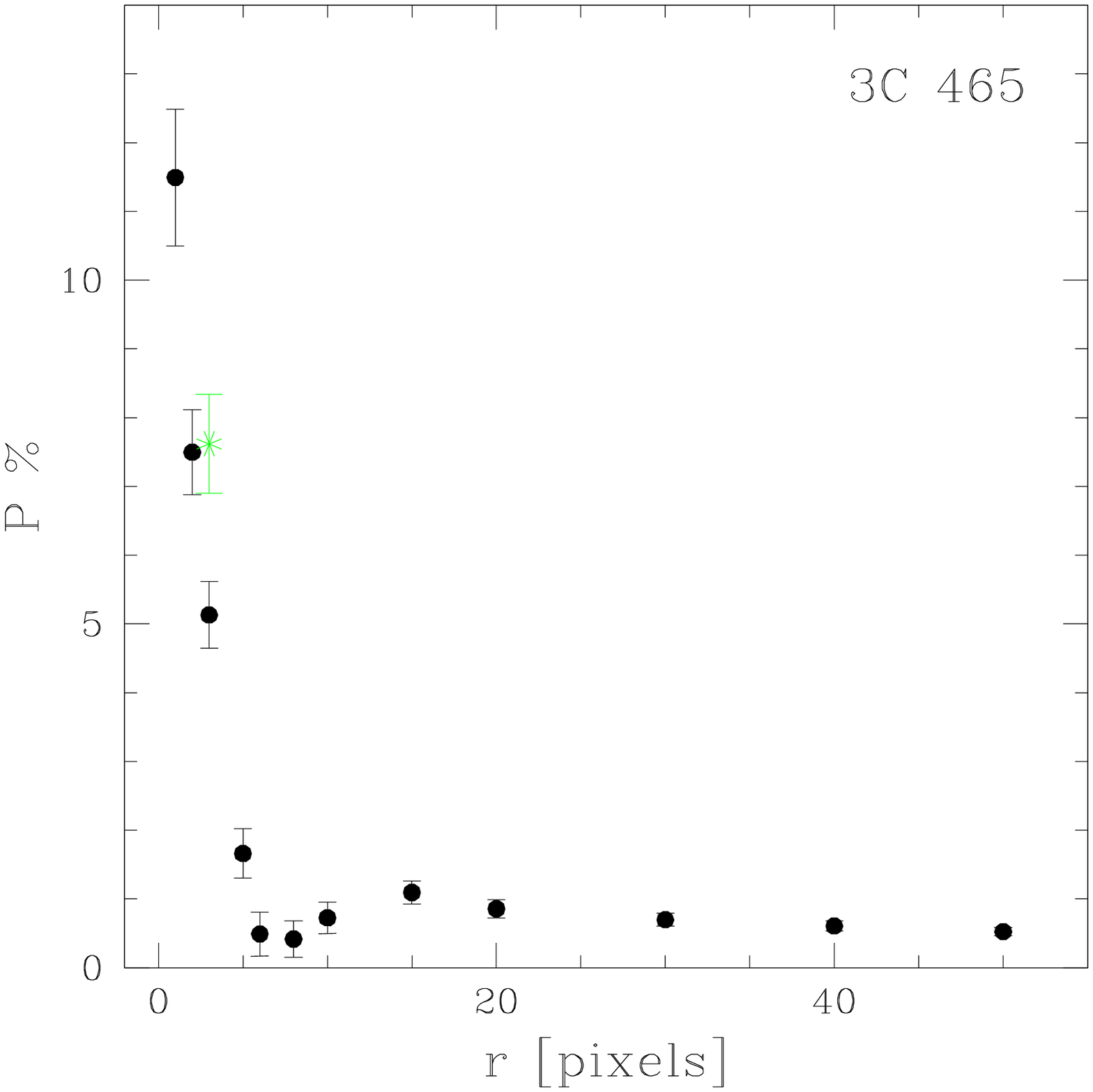,width=0.33\linewidth}}
\caption{\label{pol1} Polarization measured through apertures
of increasing radius $r$. The star at r=3 pixels
marks the polarization level derived after correcting for the
galaxy's dilution.}
\end{figure*}

In order to estimate the nuclear polarization of our targets
we obtained photometry of each galaxy through synthetic apertures
of increasing radii, up to 50 pixels, i.e. $\sim$ 1\farcs3,
centered at the galaxy's nucleus. 
In each aperture we measured the count rates in the 3 polarizers.
These were scaled according to the prescription derived in 
the Appendix \ref{calibration} and then combined to derive
the Stokes parameters. 
In Fig. \ref{pol1} we show 
the behaviour of the polarization parameters within each aperture
for all 9 radio-galaxies. 

The general behaviour is of a small 
percentage of polarization P within the largest
aperture, P$_{50} $
\ltsima  0.5\% (and thus consistent with a null value
given the calibration uncertainties)\footnote{with the exception 
of 3C~274 for which P$_{50} $ = 2.3\%, due to the presence of its
bright and highly polarized jet.}
and a substantial increase of P toward the center. 
The nuclear polarization measurements, P$_{3}$, are derived from an aperture
of 3 pixels (as suggested by the same analysis performed on the polarization
profiles on the standard stars) and
are reported in Table \ref{nucpol}. The percentage
of polarization ranges from 1.2 to 7.5 \%.
These results represent a clear indication for the presence of a polarized 
component localized at the nucleus of each galaxy of the sample. 
Only for 3C~449 the nuclear polarization
is marginally significant with P$_{3}$ = 1.2 $\pm$ 0.6 \%.
Apparently, there are no simple connections between the polarization 
properties (both percentage of polarization and position angle) 
of the FR~I nuclei and other quantities, such as total and core
radio-power, or orientation (derived from either core-dominance or dusty disk
inclination). 

These measurements refer to all light within the
3 pixels aperture, including the galaxy starlight. 
The presence of starlight (essentially unpolarized) causes 
a decrease with respect to the intrinsic level of polarization
of the nuclear source. 

\begin{figure}
\centerline{
\psfig{figure=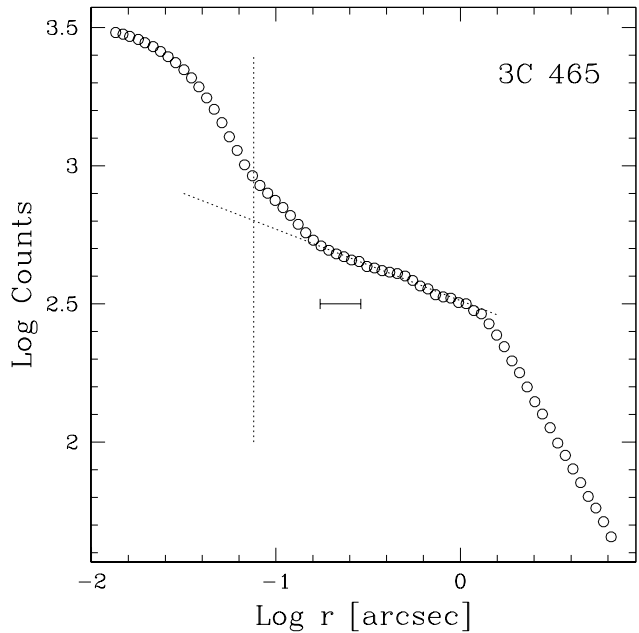,width=1.00\linewidth}
}
\caption{\label{ape} Brightness profile of 3C~465 used to
estimate the dilution from the starlight emission. 
We marked the 0\farcs1 annulus on which the galaxy's
background has been estimated (horizontal bar), 
the radius of 3 pixels
used for the nuclear aperture photometry (vertical dashed line)
and the best fitting power-law to the nuclear slope of the brightness
profile.}
\end{figure}

Different approaches can be followed
to remove the effects of this diluting component. The simplest
method is to subtract from the total intensity 
a constant galaxy's background $I_{gal}$ measured in a circum-nuclear
annulus (for which we chose a 0\farcs1\ width) 
immediately surrounding the nuclear point sources (see Fig. \ref{ape}).
The degree of polarization is then re-estimated as
$P_{\rm nuc} = P_3 \times I / (I-I_{gal})$, while the position angle
remains unchanged.
The values of $P_{\rm nuc}$ thus obtained are given in the third column
of Table \ref{nucpol}. They range from 1.9 to 11.3 \%.
In most cases the correction for dilution corresponds to an increase
of the polarization by a relatively small factor. But for the nucleus
with the lowest contrast against the galaxy's background, namely
3C~296, the correction for dilution is essential as it raises
the nuclear polarization to 11.3 \% from an uncorrected value of 2.8 \%. 

This approach neglects the increase of surface brightness toward
the center which, once modeled, would lead to a even higher 
nuclear polarization.
However, all our targets are well reproduced by a power-law
brightness profile with a small ($<$ 0.3) logarithmic slope, and we found
that this effect does not substantially affect our estimates.

\begin{table}
\caption{Nuclear polarization}
\label{nucpol}
\begin{tabular}{l c c c c c}
\hline\hline
Name & P$_3$ & P$_{\rm nuc}$ & PA & PA$_{r}$ & $\Delta$ PA\\
\hline	       			      	 	                  
3C31    &  2.5$\pm$0.4 &  3.8$\pm$0.6 &   2.3$\pm$  9.5 & -18 & 20 \\
3C66B   &  7.5$\pm$0.3 &  8.4$\pm$0.3 &  79.6$\pm$  2.3 &  50 & 30 \\
3C264   &  7.4$\pm$0.3 &  7.8$\pm$0.3 &   4.4$\pm$  2.5 &  30 & 26 \\
3C270   &  3.9$\pm$0.4 &  7.2$\pm$0.7 & 172.5$\pm$  5.6 & -93 & 86 \\
3C272.1 &  1.9$\pm$0.3 &  4.2$\pm$0.6 & 125.4$\pm$  8.1 &   1 & 56 \\
3C274   &  2.7$\pm$0.2 &  2.9$\pm$0.2 & 103.0$\pm$  4.1 & -84 &  7 \\
3C296   &  2.8$\pm$0.6 & 11.3$\pm$2.4 &  90.2$\pm$ 12.0 &  40 & 50 \\
3C449   &  1.2$\pm$0.6 &  1.9$\pm$0.9 & 115.3$\pm$ 27.9 &  11 & 76 \\
3C465   &  5.1$\pm$0.5 &  7.6$\pm$0.7 & 138.2$\pm$  5.4 & -49 &  7 \\
\hline
\end{tabular}
\\ 

P$_3$: polarization measured through a 3 pixels radius aperture;
P$_{\rm nuc}$: nuclear polarization corrected for dilution of the 
galaxy's starlight. PA$_{r}$: radio jet position angle.
\end{table}

A more subtle point is that so far we treated the starlight background as 
unpolarized. This is in general a good assumption, 
as shown by the low level of polarization at large radii shown
by Fig. \ref{pol1} which is consistent with a null value.
Nonetheless, the presence of dusty disks (in all
but one of our sources) suggests that dichroic polarization might be
produced by the passage of light through these dust structures
(we will return to this issue in Sect. \ref{off}).
This might affect particularly the nuclei with the lowest polarization
level and cause also a change in the polarization position angle.
If the galaxy is polarized, the
background must be subtracted from the nuclear aperture in the
Stokes parameters space. We then estimated also the Stokes parameter
Q and U within the region used to measure the galaxy's emission
and we subtracted a polarized background. Only marginal differences
are found with respect to the case of unpolarized background with
changes in position angles always smaller than the statistical uncertainties.

We conclude that while the dilution from the galaxy starlight is quite
significant in several galaxies, 
the final value for the nuclear polarization is rather
robust against the various methods used to treat this effect.

We also note that an additional cause of dilution 
is the presence of (unpolarized) narrow-line emission.
The presence of a Compact Emission Line Region (CELR), 
unresolved  at  0\farcs1  resolution, was shown  
by HST narrow-band imaging to be a characteristic
feature of these low-luminosity radio-galaxies \citep{capetti:cccriga}.
Measurement of the luminosity of the
H$\alpha$+NII lines (included in the spectral range covered by the F606W
filter) originating from the CELR  
is available for 8 galaxies discussed in this work, with the exception
of 3C~296.  
The average equivalent width of these lines, measured
against the nuclear sources, is $\sim$ 500 \AA. This is substantially
smaller than the F606W filter width ($\sim$ 2300 \AA) and this guarantees 
that the nuclear sources seen with the F606W filter 
are dominated by continuum emission. 
We prefer not to correct the nuclear polarization on an individual
basis, given the many sources of uncertainties related to this process,
but this implies that the genuine nuclear polarization 
should be further increased to include also this diluting component.

\section{On the off-nuclear polarization}
\label{off}

\subsection{Polarization and dust structures}
\label{dust}

\begin{figure*}
\centerline{
\psfig{figure=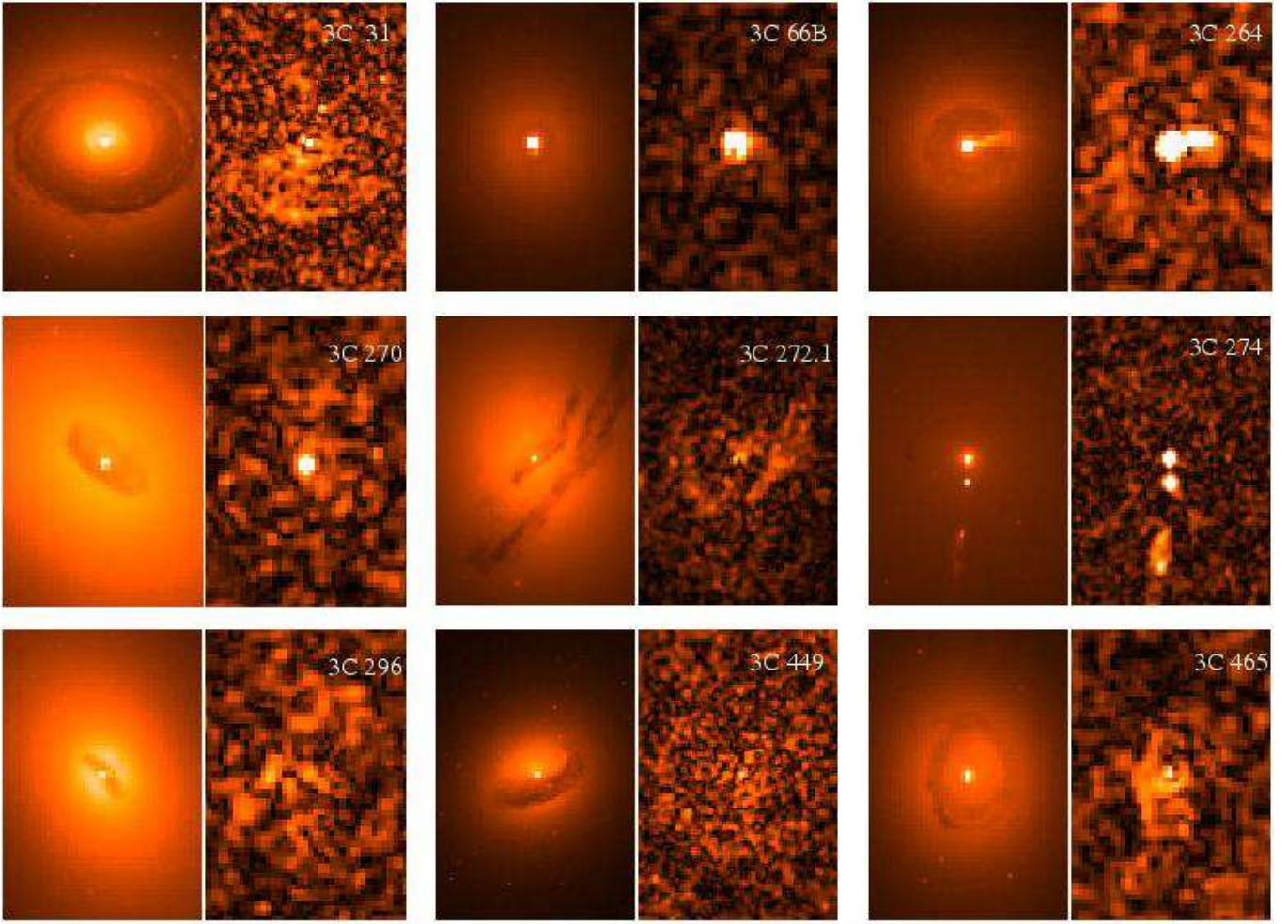,width=1.00\linewidth}
}
\caption{\label{imapol} Images of total intensity and polarized flux
for the 9 radio-galaxies.}
\end{figure*}

In Fig. \ref{imapol} we show the images of total intensity
and of polarized flux for the 9 targets, obtained after a re-binning 
of 4 $\times$ 4 pixels, approximately to 0\farcs1 $\times$ 0\farcs1.
In several cases regions of extended polarized emission 
(even leaving aside the 
polarized emission from the jets of 3C~264 a 3C~274) are detected,
always associated with dust features, thus 
suggestive of dichroic polarization.
The clearest cases are i) 3C~31 where polarization at a level
of 2 - 4 \% is seen over a fan-shaped region corresponding the
South-West side of its circumnuclear dusty disk, 
ii) 3C~465 in which the Southern side of its dusty ring 
shows a polarization of 1-2\% and iii) 3C~272.1, with 
P $\sim$ 1 - 1.5 \% over the filamentary dust structures East of the
nucleus. In all cases the polarization 
is associated to the regions of higher absorption within the
dusty structures.
Marginal polarization, below a 1\% level, might be present
over the regions of higher absorption also in 3C~296 and 3C~449.

\begin{figure}
\centerline{
\psfig{figure=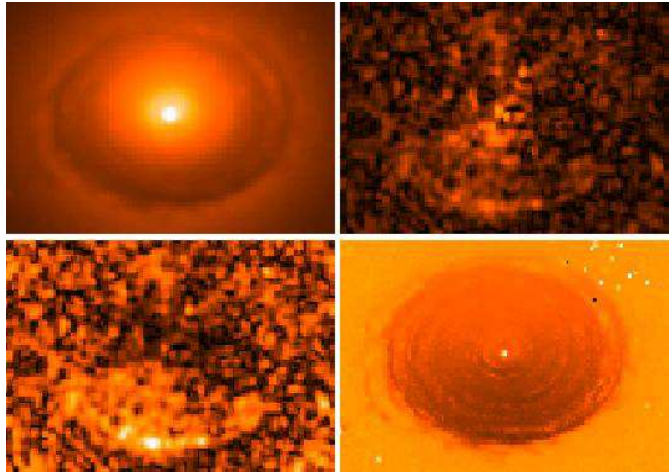,width=1.00\linewidth}
}
\caption{\label{imapol31} Images of total intensity (top left), 
polarized flux (top right),
percentage of polarization (bottom left),
and extinction for 3C~31 (bottom right).}
\end{figure}

To put the connection between
extinction and polarization on a firmer and quantitative ground
in Fig. \ref{imapol31} are shown the 
images of total intensity, polarized flux,
percentage of polarization, and extinction for 3C~31. 
The extinction map was obtained dividing the HRC images by a
NICMOS H-band (F160W) image. This enables us to
estimate the efficiency of the dichroic polarization. 
We find that, on average, a percentage of polarization of $\sim$ 2\%
corresponds to a color excess of E(B-V) $\sim$ 0.3 having adopted
the extinction law by \citet{cardelli89}. 
The polarization efficiency is thus of 
P/A$_{\rm V} \sim$ 2 \% mag$^{-1}$
(referred now to the V band).
Slightly smaller estimates, P/A$_{\rm V} \sim$ 1.5 \% mag$^{-1}$, 
are obtained for 3C~272.1 and 3C~465. These values should 
probably be considered as upper bounds since they were measured on
the objects with the highest polarization. Furthermore, 
regions of relatively high extinction but with low polarization 
are also seen. 
Overall, these estimates compare favorably with the
efficiency of dust interstellar polarization
in the Galaxy, P$_{\rm max}$ /A$_{\rm V} \sim$ 0.5 - 3 \% mag$^{-1}$
\citep{serko75}, and the
P/A$_{\rm V} \sim$ 1 \% mag$^{-1}$ 
measured in the dust lane of Cen A 
\citep{hough87,schreier96}; however a more detailed analysis
of the polarization efficiency
and a comparison with other measurements requires
multiband polarimetry as to determine its dependence with
wavelength.

\begin{figure}
\centerline{
\psfig{figure=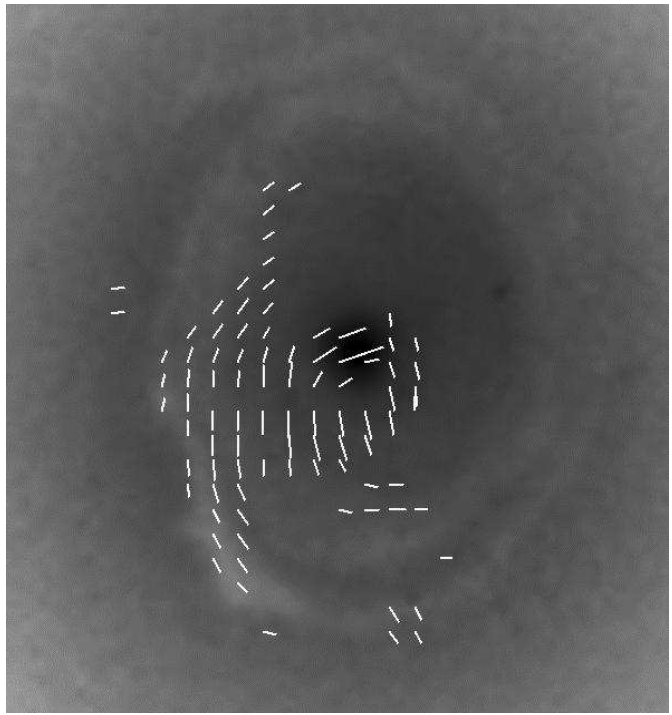,width=1.00\linewidth}
}
\caption{
\label{polvec} Polarization vectors for 3C~465 drawn at 0\farcs1\ resolution
in the regions where P $>$ 1 \%, superposed to the total intensity image.}
\end{figure}

Considering the polarization vectors,  
in 3C~31 and 3C~465 
their orientation is tangential with respect to the dusty disk
(see Fig. \ref{polvec} for the case of 3C~465).
This is what is expected when the magnetic
field responsible for the grains alignment is stretched by differential
rotation within the disk, leading to a toroidal
magnetic configuration. In 3C~272.1 
the polarization vectors are instead preferentially aligned 
with to the dust structures. This apparently different behavior 
might still be consistent with a tangential configuration
of the magnetic field if the dust features are actually part of 
a dusty disk seen close to edge-on.

We cannot exclude the possibility that the observed regions of extended
polarized
emission are due instead to scattered nuclear light, an idea that is also
supported by the orientation of the polarization vectors.
However, this interpretation appears less likely in view of the good
spatial correspondence between extinction and polarization. 
In the case of scattering in a uniform circumnuclear disk, 
we would expect the polarized regions to trace
the illumination pattern of the nuclear emission and, in case of an
anisotropic radiation field, to be possibly
enhanced along the direction of the radio-jets. This is not
what is observed.

\subsection{Are there scattered nuclear radiation beams?}
\label{beam}

Imaging polarimetry provides a tool 
to determine the geometry of the nuclear radiation field by 
measuring the large scale distribution of polarized light.
We successfully used this technique for Seyfert 
galaxies \citep{ngc1068pol,ngc5728}: around their nuclei
we detected the characteristic centro-symmetric polarization pattern
which arises from the scattering from a central source. 
The asymmetric distribution of polarized light enabled us
to determine the angular extension of the region illuminated by the nucleus
and to map the anisotropy of the nuclear radiation field.

This method is well suited in the case of FR~I 
radio-galaxies. In fact, in the frame of the unification with BL Lac
objects, they should harbour a collimated radiation beam 
which, unlike the case of BL Lac objects, is not pointing toward us.
This beam might manifest itself through scattering,
producing a collimated region of scattered light in the polarization images.
Indeed HST imaging polarimetry of the FR~I radio-galaxy Cen A 
\citep{capetti:cena} showed the presence of 
scattered nuclear radiation over an angular region 
which extends over $\sim 70^{\circ}$, centered on the radio-axis.
If due to intrinsic anisotropy,
and not to obscuring circumnuclear material, this implies the presence
 a less collimated nuclear radiation field
than expected from a mis-directed BL Lac beam with a large Lorentz factor.

\begin{figure}
\centerline{
\psfig{figure=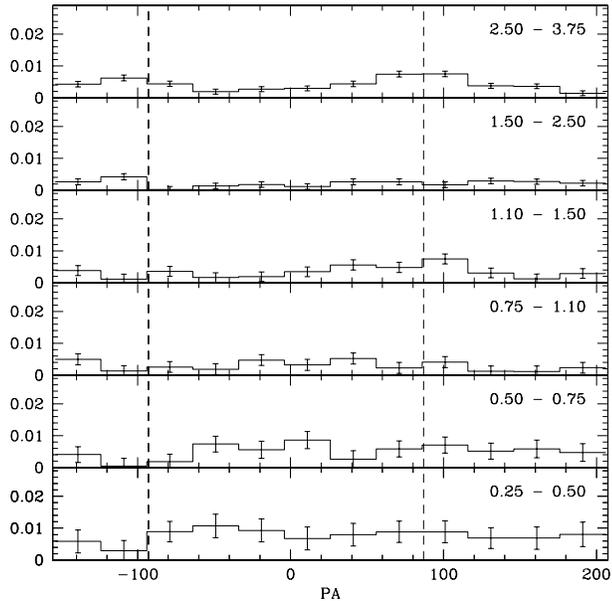,width=1.00\linewidth}
}
\caption{\label{azim} Pattern of the polarization against
the azimuthal angle (using the nucleus as origin) for 3C~270. 
The polarization has been estimated 
over wedges of 30 degrees of width over 6 different ranges of radii,
from 0\farcs25 to 3\farcs75, as reported in the top right corners of
the panels. The dashed lines mark the position angles of jet and
counter-jet.
No polarization excess is found along the radio-jets axis.}
\end{figure}

We investigated whether a similar effect is seen in our 9 targets
by exploring the azimuthal behaviour of the off-nuclear polarized
light. The signature of scattered light from a mis-oriented
beam should be an increase of polarization 
along the radio-axis, on both the jet and counter-jet side.
This effect is not seen in any of the 9 radio-galaxies
at a level of typically 0.5 \%. 
As an illustrative example, in Fig. \ref{azim} we present the
pattern of the polarization against
the azimuthal angle for 3C~270.

Unfortunately, we cannot derive quantitative
constraints from this upper limit, since 
this is the result of the interplay of several independent factors 
i.e. the intensity of the putative mis-directed beam, 
its ratio against the galaxy's starlight, and
the optical depth of the scatterers. The detection of this effect in Cen A
is probably not surprising due to the large amount of gas and dust in the
circumnuclear regions of this object that provides a larger than average
density of scatterers.

\section{Origin of nuclear polarization of low luminosity radio-galaxies.}
\label{science}

The presence of polarized light can be associated with
different physical processes:
it can be intrinsic to the radiation mechanism, as in the case of
synchrotron emission; alternatively, it can be due to the external factors,
such as scattering or dichroic transmission through dust grains.
In the following we examine which of these possibilities 
can explain the observed 
nuclear polarization in the LLRG,
taking also into account previous results obtained
from polarimetry of other classes of AGN.

\subsection{Dichroic polarization.}
\label{dichroic}

\begin{figure}
\centerline{
\psfig{figure=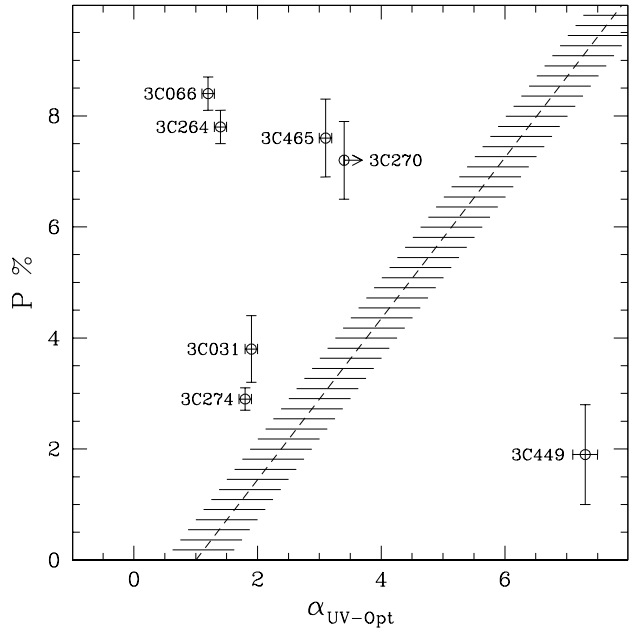,width=1.00\linewidth}
}
\caption{\label{alfa} UV to optical spectral indices against 
polarization of the FR~I nuclei. The dashed line represents 
the relationship between these two quantities adopting a dichroic
polarization efficiency of 2 \% mag$^{-1}$ and an intrinsic spectral
index of 1. The shaded region is
the locus allowed by varying the un-absorbed spectral index by $\pm$0.5.}
\end{figure}

The presence of dust features surrounding the nuclei of most of our sources
suggests that dichroic transmission might contribute to the nuclear
polarization. In Sect. \ref{off} we estimated an efficiency for the dichroic
polarization of the off-nuclear dust features of $\sim$ 2 \% mag$^{-1}$ and
this value should be considered as an upper bound since it was measured on the
regions with the highest polarization.  To account for the nuclear
polarization with dichroic transmission, at least 1 -- 6 magnitudes of
optical absorption are needed to produce the observed polarization.  The
presence of such a level of obscuration is expected to produce a significant
reddening on the nuclei of our sources.  We explored whether this is the case
by comparing the nuclear spectral indices $\alpha_{\rm{UV,O}}$ estimated in
\citet{chiaberge:uv} using UV and optical observations, available for 7 of our
sources\footnote{We also include an unpublished UV measurement for 3C~31,
obtained from an undergoing HST UV survey of the 3C sample (Prop. ID 10606, PI
Sparks), corresponding to $\alpha_{\rm{UV,O}} = 1.9$}, 
with their degree of polarization. In case of dichroic polarization,
we expect the steeper (redder) spectra to be associated with the more
polarized nuclei. This is clearly not the case. The two nuclei with available
UV data (in 3C~66B and
3C~264) with the highest polarization, $\sim$ 8\%,  have
the lowest values of $\alpha_{\rm{UV,O}}$.  Conversely, the steepest nucleus (in
3C~449) has the lowest polarization. For a more quantitative analysis, we
report in Fig. \ref{alfa} the expected relationship between $P$ and
$\alpha_{\rm{UV,O}}$, for a dichroic polarization efficiency of 2 \%
mag$^{-1}$. We also 
adopted a specific range of intrinsic spectral index,  
$\alpha_{\rm{UV,O}} = 1 \pm 0.5$ (as typical of
extragalactic synchrotron sources, 
such as BL Lacs nuclei and extended jets, e.g.
\citealt{fossati98,perlman01}) but the lack of connection between spectral index and
polarization does not depend on this assumption.

We conclude that dichroic polarization is unlikely to contribute
significantly to the nuclear polarization, unless it corresponds
to a very large range of dichroic efficiency, and 
extending to values far higher 
than measured in the dusty off-nuclear regions.

Note that the negligible level of polarization of the hosts
also rules out the possibility of a significant contribution
of interstellar Galactic polarization.

\subsection{Polarization from scattering}
\label{scattering}

\subsubsection{Case I: polar scattering from a hidden nucleus}
\label{scattering1}

The clearer examples of polarized AGN associated to
scattering are provided by Type II Seyfert (Sy 2) galaxies and powerful 
narrow-lined radio-galaxies \citep[e.g.][]{antonucci82,antonucci83,cohen99}.
Their polarized spectrum closely matches
that of a Type I (Sy 1) source, with the presence of 
a polarized featureless continuum and broad-lines,
with levels of polarization that can be as high as 30 \%.
This is interpreted as due to scattering of light from an otherwise
hidden nucleus. 
In these objects the polarization angle is generally well aligned
with the perpendicular to
the radio-axis \citep{antonucci83,ulvestad84} 
most likely the same axis of the occulting torus,
indicating that scattering takes place in a polar region located 
above the circumnuclear obscuring material.

The comparison between the position angle of the nuclear polarization
and of the radio-jets thus provides a simple test to establish whether
polar scattering might be at the origin of the observed polarization.
The orientation of the radio-axes (and their offset from 
the nuclear polarization) for our sources are
given in Table \ref{nucpol}. They are taken 
from VLBI/VLBA observations \citep{xu00,venturi95}, except 
for 3C~296 for which we used the VLA maps by \citet{hardcastle97}.

There is no preferential alignment between radio-axis and polarization,
as the whole range of possible offsets (between 0 and 90 degrees)
$\Delta PA$ is observed.
In 5 galaxies (namely 3C~31, 3C~66B, 3C~264, 3C~274 and 3C~465) 
the polarization angles are parallel to the jet axis, to within
30 degrees, while for two sources (3C~270 and 3C~449) 
the nuclear polarization is close to
perpendicular to the radio jets. For the remaining two galaxies
(3C~272.1 and 3C~296) the offset is of $\sim$ 50 degrees. 
These large misalignments from the perpendicularity expected 
in the case of polar scattering
argue against this interpretation to account for the FR~I
nuclear polarization. 

\subsubsection{Case II: equatorial scattering from a visible nucleus}
\label{scattering2}

Type I Seyfert nuclei (and radio-quiet quasars) are also
often polarized, although generally at significantly lower levels
than Sy 2; 
their polarization angle is usually 
aligned with the radio-jet, although there are examples of 
perpendicular polarization \citep{stockman79,antonucci83}. 
This has been interpreted
as a contribution to the polarized flux of scattered light 
from an equatorial disk-like region,
which is hidden in the case of Sy 2 \citep{smith04}. The relative weight
of the polar vs. equatorial scattering sets the dominance of
polarization perpendicular or parallel to the AGN symmetry axis.
Thus, in the case of scattering from a visible nucleus,
there are no simple predictions for the orientation of the polarization.
The analysis of a sample of 36 Type I Seyfert
\citep{smith02} gives a median
polarization of 0.5 \%, with only one object above a level of 2 \%.
The significantly lower polarization level in Sy 1,
with respect to type 2, is naturally accounted
since in this situation the light from the unocculted nucleus 
provides a copious source of diluting unpolarized light. 

\begin{figure}
\centerline{
\psfig{figure=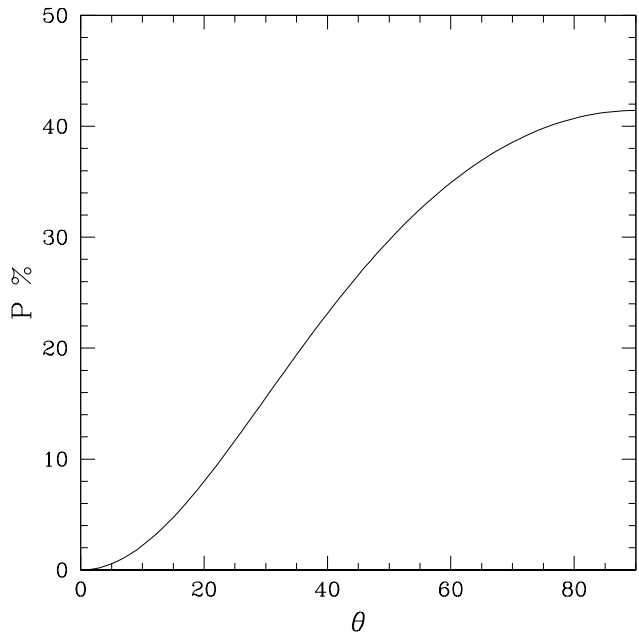,width=1.00\linewidth}
}
\caption{\label{eq} Dependence of the 
degree of polarization of light scattered
in an optically (and geometrically) thin circumnuclear disk on the
orientation of the disk axis with respect to the line of sight.
90 degrees corresponds to an edge-on disk.}
\end{figure}

More quantitatively the level of nuclear polarization is set by:
i) the optical depth $\tau$ of the scattering region and
ii)  the covering angle  of the  scattering material $\Omega$ (that together determine
the fraction of scattered nuclear light); but the observed 
degree of polarization depends also on the geometry of
the system i.e. iii) on the scattering  angle and iv) on the geometrical
cancellation produced by summing over polarization vectors
with different orientation within the scattering region.
In Fig. \ref{eq} we show the dependence on orientation of the 
fractional polarization of the scattered
light in a thin disk.
The optimal geometrical setting to produce a high polarization
is an edge-on thin disk \citep{smith02} that corresponds to a degree of 
polarization of $ f \sim$ 40 \% (while the average for a randomly
oriented sample is $ f \sim$ 30 \%). 
The nuclear polarization level is then given by
$ P =  f \Omega (1 - e^{-\tau}) $.
Our observations provided measurements in the range $\sim$ 2 - 11
\% with a median value of 7 \%, more than an order of magnitude 
higher than observed in Seyfert 1. 
In order to achieve this level of polarization an implausible
high fraction of nuclear light ($\Omega (1-e^{-\tau}) \gta 0.2 $) is needed
to be scattered into our line of sight.

\subsection{Synchrotron emission}
\label{synchro}

Synchrotron emission is intrinsically polarized and it 
has been firmly established that it represents the dominant
emission process associated to several classes of AGN. 
In particular this is the case of all objects falling within the
definition of Blazars, including BL Lac objects and core-dominated
radio-loud quasars, that are indeed highly polarized
\citep[e.g.][]{angel80}. 

In the framework of the most promising unifying model for 
weak radio--loud AGN, LLRG are identified
as the parent (misdirected) population of BL Lac objects
(e.g. \citealt{urry95} for a review). The non--thermal continuum
emission of BL Lacs 
results from the observation of plasma moving at relativistic speed
at a small angle with respect to its direction of motion (Blandford \&
Rees 1978).  This emission is expected to be present also in radio-galaxies 
although not amplified by relativistic beaming, or even de-amplified.

Here we then compare the polarization properties of LLRG and BL Lacs.
Such a comparison is complicated by the high variability
of BL Lacs polarization and its dependence on wavelength
\citep{brindle86}, by
the effects of different viewing angle and relativistic aberration. 
Nonetheless, we find that 
the degree of polarization of BL Lac objects is quantitatively
similar to the measurements presented here for LLRG: for
example, the median value of polarization
in the sample of 20 radio-selected BL Lacs studied by \citet{impey90}
is $\sim 8 \%$, while
X-ray selected BL Lacs are apparently slightly 
less polarized \citep{jannuzi94}.
Concerning the polarization angle, \citet{gabuzda06} showed 
that BL Lac have a close  
alignment between radio and optical polarization; conversely, although 
the radio polarization is preferentially aligned
with the jet direction (within 20 degrees), a smaller but substantial
population shows polarization orthogonal to the jet axis (within 30 degrees).
Again, this is analogous to the situation found for LLRG.

We conclude that the polarization properties of LLRG nuclei closely
resemble those of BL Lac objects, strongly favoring the
interpretation that they are dominated by synchrotron emission.

The direct comparison between the optical and radio core
polarization properties 
of our radio-galaxies is in general hampered by the lack of high quality
radio data. Nonetheless, 
\citet{kharb05} provide VLBI measurements of the radio core polarization
of 3 galaxies of our sample (namely 3C~66B, 3C~264 and 3C~270). 
In all 3 cases the percentage of polarization is quite low (between
0.4 and 1.1 \%) and much lower than measured in the optical
where we measured typically a 8 \% level. This is in line
with the result we obtained for Cen A \citep{capetti:cena}, 
where a 11 \% near infrared nuclear polarization is contrasted by
a 0.46 \% of radio core polarization \citep{burns83}.

The most likely explanation for the different polarization
level of the radio and optical nuclei is the presence of
a depolarizing medium located in front of the active nucleus.
Evidence for circumnuclear gas in FR~I nuclei came recently from
X-ray observations of these objects \citep{balmaverde06}; 
we detected  nuclear X-ray
absorption in 4 galaxies at a level of $N_H \sim (0.2-5)\,\, 10^{22}$  cm$^{-2}$.
X-ray absorbed sources are associated with the presence
of highly inclined dusty disks seen in the HST images.  
This suggests the existence of a flattened X-ray absorber,
but of much lower optical depth than in classical obscuring tori.
Due to the $\lambda^2$ dependence of Faraday rotation, this gas
might act as a depolarizing screen onto the nucleus in the radio band,
while leaving unaltered the optical polarization level. 

\subsection{Polarized emission 
from radiatively  inefficient accretion flows}
\label{riaf}

In the introduction we discussed the possibility that the faint
optical nuclei detected in FR I radio-galaxies might originate from a
low radiative efficency accretion flow. Is this consistent with our
polarization measures?  Indeed, recent observations indicate that also
the emission associated with the black hole at the center of our own
Galaxy - a likely candidate system for such accretion regime -
produces polarized light outside the radio-band.  In fact Sgr A$^*$
showed high polarization (up to 40 \%) associated to flares in the K
band \citep{trippe06}.  This might be taken as an indication that
accretion in such conditions is able to produce significant
polarization, opening an alternative explanation to the association
with the jet emission we proposed.

The low-frequency emission expected from such flows originates as
cyclo-synchrotron emission from thermal plasma, which is expected to
be polarized.  Given the temperature, density and magnetic energy
density conditions theoretically inferred, this spectral
component should dominate - at least in the simplest scenario - at,
typically, mm wavelengths \citep[e.g.][]{mahadevan97}.

Nonetheless, depending on the physical parameters describing a
RIAF, the polarized cyclo-synchrotron radiation might still dominate,
over Comptonization and bremsstrahlung emission, also at higher
frequencies. While no unique prediction is possible, in this
situation, the emission in the optical band would be typically
characterized by a steep spectrum.  This possibility is then
disfavored by the flat spectral energy distribution of FR~I nuclei in
the UV-optical band found by \citet{chiaberge:uv}.

The situation could be more complicate. The very presence of
rapidly varying polarized IR (and X-ray) emission in Sgr A$^*$ have
led to the hypothesis that flaring is associated to non-thermal
synchrotron radiation, from a dissipation region either in the jet or
within the accreting flow. Apparently, this also leaves the
possibility of RIAF emission open also for FR~I.  However, it should
be stressed that while high levels of polarization in Sgr A$^*$ are
observed during flares, the detection of nuclear polarization in
8 out 9 of our FR~I sources (leaving aside the 2 $\sigma$ detection in
3C~449) argues for a more stable process.

\section{Summary and discussion}
\label{summary}

We obtained optical imaging polarimetry with the ACS/HRC
aboard the HST of the 9 closest radio-galaxies in the 3C catalogue with an FR~I
morphology. The analysis of the observations of 
standard polarization targets lead to a calibration of the
ACS/HRC polarimeter accurate to a level of 0.4\%.

The nuclear sources seen in direct HST images in these galaxies are found to
be polarized with levels in the range $\sim$ 2 - 11 \% with a median value of
7 \%.  There is no clear preferential alignment between radio-axis and
polarization vectors; 
the whole range of possible offsets is observed, with only a
mild preference (in 5 objects) for polarization parallel to within 30 degrees
to the radio-jets, but cases of perpendicular polarization are also found,
similar to the case of the nuclear polarization of Cen A.

In order to explore the origin of the nuclear polarization, we discussed the
different mechanisms that produce polarized emission and conclude that the
most likely interpretation is a synchrotron origin for the optical nuclei.  In
fact, polarization induced by dichroic transmission predicts that the most
polarized nuclei should be the most absorbed, while their UV-optical spectral
slope indicates that this is not the case.  The possibility of polarization
produced by polar scattering of light from an obscured nucleus (similarly to
what is observed in Type II Seyfert and narrow-line radio-galaxies) contrasts
with the lack of perpendicularity between radio-jets and nuclear polarization.
Scattering occurs also in Type I AGN, but in this case the unocculted nucleus
provides a copious source of unpolarized light, and indeed their typical
polarization is of only $\sim$ 0.5 \%.  In order to achieve the level of
polarization measured in LLRG an implausible high fraction of nuclear light
(\gta 0.2) should be scattered into our line of sight.

This leaves us with the sole possibility that the nuclear light in FR~I is
{\sl intrinsically} polarized, such as in the case of synchrotron emission.
This idea is strengthened by the analogy with the polarization properties of
BL Lac objects (from the point of view of percentage of polarization and of
the complex relationship between polarization position angle and radio-jet
axis) where the dominant contribution of synchrotron light is well
established.  Incidentally, the similarity of LLRG nuclei and BL Lacs
polarization adds to the list of properties shared by these two classes of
objects, strengthening the case in favour of the FR~I/BL Lac unified model.

Polarized emission is also expected in case of
low radiative efficency accretion flows, although restricted
to the spectral region dominated by cyclo-synchrotron emission.
It is possible to ascribe the optical polarization to this process
only if the cyclo-synchrotron component, peaking in the mm
region, remains the dominant radiation process into the optical band.
While this cannot be excluded on purely theoretical ground, 
in such case the optical light would represent 
an exponential emission tail and it should show a very steep spectrum,
contrarily to what is seen in FR~I nuclei.
Similarly, the presence of rapidly varying polarized IR 
emission in Sgr A have lead to the hypothesis that (at least) flaring is
associated to non-thermal synchrotron radiation.  
However, it should be stressed
that polarization in Sgr A$^*$ is only observed during flares, while the
detection of nuclear polarization in the FR~I sources
argues for a more stable process.

These results, taken together, lead us to the conclusion that the optical
emission of FR~I nuclei is most likely
dominated by synchrotron emission from the base of their jets.

This conclusion relies on the absence of an additional nuclear component,
hidden to our view, 
other and (if it has to alter considerably our findings) substantially
brighter than the synchrotron nuclei.
The results presented here tell us that we are not seeing
compact scattering regions illuminated by a bright nucleus,
but the existence of a bright, hidden, nuclear source can not be ruled out on
the base of the present data alone. 
 
However, this possibility 
sets the following
requirements on the nuclear structure of these radio-galaxies: i) the putative
extra component must be hidden to our view in the optical band; since our
sample is unbiased with respect to orientation, the occulting material should
essentially take the form of a sphere, with, necessarily, only a free path
coincident with the radio-jets, ii) it must be hidden also in the X-ray band;
\citet{evans06} explored this possibility and concluded that this is
incompatible with X-ray observations of FR~I sources unless they are covered
by extreme gas column densities, with $N_{\rm H} > 10^{24}$ cm$^{-2}$,
far higher than in FR~II radio-galaxies; iii)
the optical depth toward the true nucleus must be also extremely high in the
mid-infrared to prevent the view of the nuclear emission reprocessed by the
circumnuclear dust; in fact 3C~274 does not show a substantial MIR excess with
respect to the extrapolation of the synchrotron component to this band
\citep{whysong04} and Spitzer observations of another 4 objects of our sample
indicates an overall low activity level \citep{ogle06}, in line
with previous findings based on ISO data \citep{muller04}; 
iv) the correlation
between the optical nuclear continuum and the luminosity of the emission lines
in these radio-galaxies suggests that the Narrow Line Region gas is
ionized by the jet emission \citep{capetti:cccriga}; apparently, also the NLR
gas does not see any additional ionizing photon field;
v) since the jet component is visible in both optical and X-ray bands,
it must originate outside the obscuring structure. 
We conclude that, most
likely, the non-thermal emission represents the dominant
radiative manifestation in the FR~I nuclei.

The association between optical nuclei and jet emission was already suggested
by \citet{chiaberge:ccc} and \citet{balmaverde06} based on the close
correlation between radio and optical nuclear luminosities and it is fully
supported
by the present polarimetric analysis.  The radio/optical/X-ray correlations
recently found by \citet{paper2} in a sample of radio-loud early-type galaxies
extend this connection toward lower power, down to radio luminosities $\nu
$L$_r \sim 10^{36}$ erg/s and to optical luminosities as low as $\sim 10^{38}$
erg/s. The reasonable (and testable) extension of these polarimetric results
to this latter sample would set a limit to any emission from the accretion
process as low as L/L$_{\rm Edd} $\ltsima$ 10^{-9}$ in the optical and
X-ray bands. In contrast, as already discussed in \citet{paper3}, their Bondi
accretion rates, $\dot{M}_B$, derived from Chandra observations
\citep[e.g.][]{pellegrini05} are relatively large, $\dot{M}_B/\dot{M}_{\rm
  Edd} \sim 10^{-2} - 10^{-4}$.  This raises the possibility that most of the
accreting gas fails to reach the central object, possibly carried away by
outflows originating at relatively large radii.  These findings represent a
challenge to the models of jet production, as they require that the energy
released by gas accretion onto the supermassive black-hole is channeled with
high efficiency in jet mechanical energy while leaving the accretion disk in a
extremely low radiative state.

In addition to the polarization in the nuclei and from the large scale optical
jets (which are presented elsewhere, \citealt{perlman06}), regions of extended
polarized emission are also detected. They are closely cospatial with dust
features (mainly taking the form of dusty circumnuclear disks) indicative of
polarization due to dichroic transmission. In the two clearest examples the
polarization vectors are tangential to the disks as expected when the magnetic
field responsible for the grains alignment is stretched by differential
rotation.

Conversely, we fail to detect an extended component 
of polarized light aligned with the
radio-jet axis. This component might have been expected in the frame of the
unification of FR~I with BL Lac objects, since FR~I should harbour a
collimated radiation beam misaligned with respect to our line of sight that
might be visible in scattered light. Indeed, this effect is present in the
radio-galaxy Cen A that shows the characteristic
centro-symmetric polarization pattern centered on the
radio-axis. We did not find the polarimetric signature of this effect in any
of the galaxies of the
present sample, with limits to the excess of polarized light along the
radio-axis at a level \ltsima 0.5\%. 
We cannot derive quantitative
constraints from this upper limit as it depends on the intensity of the
putative mis-directed beam (with respect to the galaxy's starlight) but also
on the optical depth of the scatterers. 
Thus, our the results are inconclusive on whether 
a radiation misaligned beam is present in FR~I.

\appendix

\section{The ACS/HRC polarimeter}
\label{calibration}

Several issues concerning the use of the HRC as a polarimeter
must be addressed before deriving the polarization properties of
scientific targets, in particular related to the possible
presence of instrumental polarization. They have been discussed in detail in
\citet{biretta04a} and are here only briefly summarized.  
Instrumental polarization might arise due to the presence of 
mirrors with large incident angles and of strongly tilted CCD detectors.
The first point is particularly important since light reflected by 
mirrors may be affected by both di-attenuation 
(instrumental polarization caused by the different reflectivity for
light polarized parallel or perpendicular to the incidence angle) 
and phase-retardance (the conversion of linear into elliptical
polarization).
These result in an instrumental polarization that varies with 
the telescope orientation as well as with the polarization state of
the incident light. In addition, the polarization measurements are
affected by i) the accuracy of the flat-field in the individual
polarizers, ii) the polarizers orientation, which might differ from the nominal
value and iii) by the polarizers relative transmission.

\begin{table*}
\caption{Standard stars observations.}
\label{standard}
\centering
\begin{tabular}{l c c c c c c c c c}
\hline\hline
Star & comment& & \multicolumn{3}{c}{Count rates} & \multicolumn{2}{c}{Method I} &\multicolumn{2}{c}{Method II}   \\
     &        & PAV3& POL0V & POL60V & POL120V  & P \% & PA & P \% & PA \\
\hline	       			      	 	                  
GD319  & unpol.&    318  &  58906 &  64230 &  65150 & 0.2$\pm$0.6 &      --       & 0.2$\pm$0.6 &      --\\
GD319  & unpol.&    276  &  58806 &  64561 &  65300 & 0.2$\pm$0.6 &      --       & 0.6$\pm$0.6 &      --\\
G191B2B& unpol.&     91  & 138556 & 151884 & 153381 & 0.1$\pm$0.4 &      --       & 0.5$\pm$0.4 &      --\\
BD+64D106  & pol.  &  89 & 607846 & 723703 & 709541 & 4.9$\pm$0.2 &  97.3$\pm$2.1 & 4.3$\pm$0.2 & 106.0$\pm$2.4 \\
BD+64D106  & pol.  & 135 & 651915 & 716861 & 662958 & 5.8$\pm$0.2 &  97.1$\pm$1.7 & 4.6$\pm$0.2 &  93.7$\pm$2.2 \\
BD+64D106  & pol.  & -76 & 647274 & 760131 & 721303 & 4.6$\pm$0.2 &  96.2$\pm$2.2 & 3.4$\pm$0.2 & 102.1$\pm$3.0 \\
BD+64D106  & pol.  & -76 & 645337 & 760477 & 719661 & 4.8$\pm$0.2 &  96.3$\pm$2.1 & 3.6$\pm$0.2 & 101.9$\pm$2.8 \\
BD+64D106  & pol.  & -76 & 649304 & 769996 & 730473 & 5.0$\pm$0.2 &  99.1$\pm$2.0 & 3.9$\pm$0.2 & 105.1$\pm$2.6 \\
BD+64D106  & pol.  & -16 & 704965 & 718297 & 713385 & 5.6$\pm$0.2 & 100.1$\pm$1.7 & 5.4$\pm$0.2 &  92.5$\pm$1.8 \\
BD+64D106  & pol.  &  44 & 666275 & 726582 & 790493 & 4.7$\pm$0.2 &  95.8$\pm$2.1 & 6.2$\pm$0.2 &  99.5$\pm$1.6 \\
BD+64D106  & pol.  &  44 & 663241 & 720569 & 791937 & 5.3$\pm$0.2 &  96.8$\pm$1.9 & 6.8$\pm$0.2 & 100.0$\pm$1.5 \\
\hline
BD+64D106  & \multicolumn{5}{c}{Averaged values / rms } & 5.1 / 0.4 & 97.3 / 1.5
& 4.8 / 1.2 & 100.1 / 4.9 \\
\end{tabular}
\\ 
\end{table*}

Here we present the results obtained concerning the presence
of instrumental polarization based on the analysis
of the calibration observations of standard polarimetric stars 
limiting to our filter/instrument combination i.e. HRC+F606W. 

Two unpolarized standard stars, GD319 and G191B2B, lead to measurements
of the count rates through each of the 3 polarizers 
(using an aperture of 50 pixels), see Table \ref{standard}. 
They indicate a substantial departure from an ideal polarimeter,
for which one would expect equal count rates. 
Conversely, these observations lead to a measurement of the relative ratios
of POL60V/POL0V = 1.095 $\pm$ 0.004 and POL120V/POL0V = 1.109 $\pm$
0.004, where we used
the average values from the 3 observations (GD319 was observed twice).
The individual ratios measurements differ by less than 0.3\%, a level
similar to the statistical uncertainties.

We then estimated the count rates for the available
observations of a polarized standard star, BD+64D106.
It was observed at 5 different epochs and, most importantly for our discussion,
at 5 different telescope roll angles. We again derived the count rates
which are presented in Table \ref{standard}. 
Different exposures were not combined,
but treated individually, for a total of 8 datasets.

We followed two methods to derive the polarization parameters of BD+64D106
from these observations. In the first approach we ascribed the different count
rates measured for the unpolarized standards to a difference in the
transmission of the 3 polarizers. We then simply scaled the count rates for
BD+64D106, by 1.095 and 1.109 in the POL60V and POL120V images respectively,
and only then we estimated the Stokes parameters.  In Table \ref{standard} we
present the resulting values of P and PA.  The averaged values are P =
5.1$\pm$0.4 \% and PA = 97.3$\pm$1.5 (the errors quoted are the rms of the
individual measurements) in very close agreement with the ground based
measurements of this star in the R band, P = 5.150 $\pm$ 0.098 at PA
96.74$\pm$0.54 \citep{schmidt92}.

In the second method we corrected all measurements ascribing the count rates
differences to the presence of instrumental polarization.  The count rates
measured for the unpolarized stars correspond to an instrumental polarization
with Stokes parameter Q=-6.36 \% and U=0.76 \% (in the instrument reference
frame).  We estimated the 'raw' Stokes parameters, then removed the
instrumental polarization in the Stokes parameter space from all measurements.
The resulting polarization parameters, P = 4.8$\pm$1.2 \% and PA =
100.1$\pm$4.9, are still in a reasonable agreement with the known polarization
of BD+64D106, but they do not match as closely as those derived from the first
method.  Furthermore, they show a larger dispersion as well as substantial
changes related to the telescope roll angle, unlike what we found from the
first method.

But  a   better  insight  into  these  issues   comes  from  combining
opportunely the observations obtained at PAV3 = -76, -16 and 44. These
observations were planned to be separated by exactly 60 degrees in the
telescope  roll angle. Combining  the images  taken through  one given
polarizer at the three different  roll angles, it is equivalent to the
situation of  observing the target through  three polarizers separated
by 60 degrees  at the same roll angle. 
It is also worth noting that by rolling the telescope one rotates any
instrumental polarization in the Equatorial system enabling us
to determine it from observations of standard stars at different rolls.
It is  thus possible to measure
the  polarization  parameters, removing  the  uncertainty  due to  the
filter  transmission.   In particular,  the  ratio  between the  total
intensities   is   a   direct   measure   of   the   relative   filter
transmission. We obtained  I60/I0 = 1.092$\pm$ 0.002 
and I120/I0  = 1.102$\pm$ 0.002, having
indicated with I0, I60 and  I120 the total intensity derived combining
the observations  at three  roll angles of  POL0V, POL60V  and POL120V
respectively.  These values  differ  only marginally,  and within  the
statistical  uncertainties,   from  the  ratios   estimated  from  the
unpolarized stars, i.e. 1.095 and 1.109.

This strongly supports the interpretation that the dominant
contribution to the observed polarization level for the
unpolarized standard stars comes from differences in the filter transmission.
By applying a simple correction to the count rates we are able to
reproduce accurately the expected values for the polarized standard star.
The resulting dispersion in the polarization degree is 0.4\%,
with PA variations of less than 2 degrees. 
These values can be used as estimates for the 
uncertainties due to any residual instrumental polarization as well
as to the flat field accuracy on the polarization measurements
for our science targets (in addition to the
statistical uncertainties). 

The observations of the unpolarized standard stars can also be used
to test if there are any differences between the point spread function
produced by each polarizers. This is important since we calibrated
our polarization measurements using large apertures (50 pixels). 
However, since for our science targets we are interested in measuring
the nuclear polarization we will be forced to use substantially smaller
apertures. 
If, on a scale of a few pixels, there are differences in the
PSF of each polarizer, we might obtain
spurious polarization measurements. 

\begin{figure}
\label{stars}
\centerline{
\psfig{figure=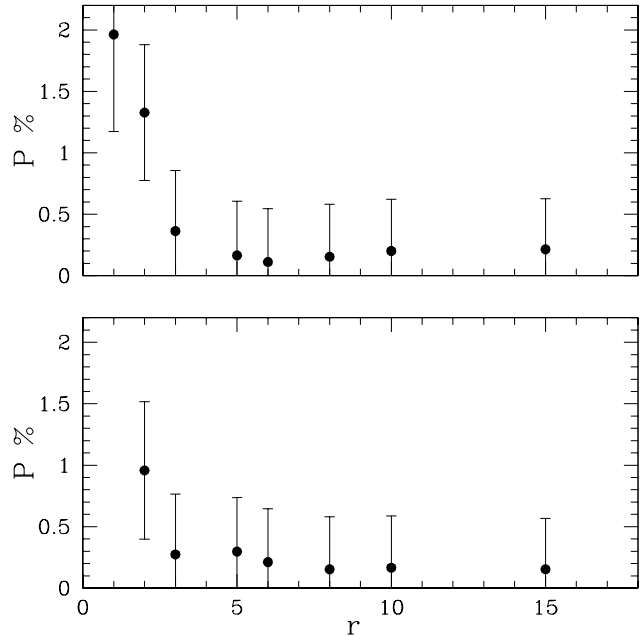,width=1.00\linewidth}}
\caption{Percentage of polarization
measured from two sets of observations of an unpolarized standard star.
Different circular synthetic apertures were used. The polarization 
level is consistent with a null level down to an aperture of 3 pixels.
Only below this radius small differences between the Point Spread Function
through the polarizers produce a spurious polarization.}
\end{figure}

To test this effect, we measured the polarization
parameters using the count rates 
integrated through several synthetic apertures, 
ranging from a radius of
1 pixel to 50 pixels. The derived levels of
polarization are shown in Fig. \ref{stars}.
The polarization is below a 0.4 \% level
down to a radius of 3 pixels. This is consistent with a null
value within the errors
due to the polarization calibration and with the statistical
uncertainties. Conversely,
using even smaller apertures, the polarization level increases 
to $\sim$ 1 \% at 2 pixels and even larger values using a 1 pixel aperture. 
Since we are dealing with unpolarized  stars, 
this is a spurious polarization signal, most likely to be ascribed
to a combination of different PSF and possibly also to small
mis-centering of the stars. 
This indicates that robust polarimetric measurements 
can be obtained with synthetic apertures of radius as small as
r $\gta$ 3 pixels, with no appreciable
differences in accuracy with respect to those used for the
polarimetric calibrations that were derived from large apertures. 

We finally note that the lack of sufficient calibration data
in the UV band prevents us from following the strategy
adopted for the optical data and from analyzing the data for 
the two radio-galaxies observed in this band. 

\section{On the bias in the polarization measurements.}
\label{bias}

It is well known that linear polarization measurements 
are affected by a bias due to the fact the $P$ is a 
definitive positive quantity. Consequently, 
it follows a Rice (and not a
Gaussian) probability
distribution \citep[e.g.][]{wardle74}. 

The resulting effect is that the degree of polarization
is overestimated at low signal-to-noise ratios. E.g., 
a null level of polarization
corresponds to an average measurement of $P = \sigma_P$.
This is clearly seen
in Fig. \ref{stars}.1 where the level of polarization
for the unpolarized stars decreases at large radii to $\sim 0.2 \%$, 
approximately at a 1 $\sigma$ level. A similar behaviour is found
in the azimuthal behaviour of the off-nuclear polarized
light (Fig. \ref{azim}) where the minimum polarization is $\sim$ 0.5 \%.
This effect becomes rapidly less important at higher signal-to-noise
ratios. In fact, the polarization overestimate 
is already reduced to a factor of $\sim$ 1.1
at $P = 2 \times \sigma_P$ 
and becomes essentially negligible at higher significance
level. 

Several methods to de-bias polarization data have been proposed
\citep[e.g.][]{simmons85} based on the probability distribution of
the errors in the Stokes parameters. However, such an approach requires that
the measurement errors are dominated by statistics. This
is not necessarily the case for our measurements in which 
a possible contribution of instrumental polarization might be present.
We therefore preferred not to attempt any correction for the polarization
bias and decided to consider in our discussion 
only regions of sufficiently high polarization,
$P > 2 \times \sigma_P$, not significantly affected by this bias.

\end{document}